\documentclass[pra,twocolumn,showpacs,amsfonts,amsmath,floatfix,superscriptaddress]{revtex4-2}

\usepackage[svgnames]{xcolor}

 \usepackage[
            colorlinks=true,        
            allcolors = black,  
            citecolor=DarkBlue, 
            linkcolor=DarkBlue,
            urlcolor=DarkBlue
        ]{hyperref}  
\usepackage[utf8]{inputenc}
\usepackage[english]{babel}
\usepackage[T1]{fontenc}
\usepackage{amsmath}
\usepackage{amssymb}
\usepackage{hyperref}
\usepackage{caption}
\usepackage{tikz}
\usepackage{lipsum}
\usepackage{graphicx}
\usepackage{epsfig}
\usepackage{epstopdf}
\usepackage{braket}
\usepackage{hyperref}
\usepackage[capitalize]{cleveref}
\usepackage{amsthm}
\usepackage{enumitem}
\usepackage{dsfont}
\usepackage{bbold}
\usepackage{bm}
\usepackage{physics}
\usepackage{eurosym}
\usepackage{diagbox}
\usepackage{array}
\usepackage{float}
\newcolumntype{C}[1]{>{\centering\arraybackslash}p{#1}}
\newcolumntype{D}{>{\centering\arraybackslash}p{1cm}}
\usepackage[ruled,vlined]{algorithm2e}
\usepackage{comment}

\newcommand{\ii}{\mathrm{i}}
\newcommand{\e}{\mathrm{e}}

\usepackage[colorinlistoftodos]{todonotes}

\begin{document}

%\title{Witnessing and quantifying non-Gaussianity in superconducting bosonic systems}

\title{Witnessing and quantifying stellar rank with the Wigner function}

\author{Oliver Hahn}
%\email{hahn@g.ecc.u-tokyo.ac.jp}
%\orcid{0000-0003-1677-8696}
\thanks{These authors contributed equally to this work.}
\affiliation{Department of Basic Science, The University of Tokyo, 3-8-1 Komaba, Meguro-ku, Tokyo, 153-8902, Japan}

\author{Alex Maltesson}
\thanks{These authors contributed equally to this work.}
\affiliation{Department of Microtechnology and Nanoscience, Chalmers University of Technology, Sweden , SE-412 96 G\"{o}teborg, Sweden}

\author{Varun Upreti}
\thanks{These authors contributed equally to this work.}
\affiliation{DIENS, \'Ecole Normale Sup\'erieure, PSL University, CNRS, INRIA, 45 rue d’Ulm, Paris, 75005, France}

\author{Marike Ellerbroek}
\affiliation{Department of Microtechnology and Nanoscience, Chalmers University of Technology, Sweden , SE-412 96 G\"{o}teborg, Sweden}

\author{Alessandro Ferraro}
\affiliation{Dipartimento di Fisica ``Aldo Pontremoli'',
Università degli Studi di Milano, I-20133 Milano, Italy}
\affiliation{Centre for Theoretical Atomic, Molecular and Optical Physics, Queen's University Belfast, Belfast BT7 1NN, United Kingdom}

\author{Saleh Rahimi-Keshari}
\affiliation{Wallenberg Centre for Quantum Technology, Department of Microtechnology and Nanoscience, Chalmers University of Technology, Sweden , SE-412 96 G\"{o}teborg, Sweden}
\affiliation{School of Quantum Physics and Matter, Institute for Research in Fundamental Sciences (IPM), P.O. Box 19395-5531, Tehran, Iran}

\author{Ulysse Chabaud}
% \email{ulysse.chabaud@inria.fr}
% %\orcid{0000-0003-0135-9819}
\affiliation{DIENS, \'Ecole Normale Sup\'erieure, PSL University, CNRS, INRIA, 45 rue d’Ulm, Paris, 75005, France}

\author{Giulia Ferrini}
\affiliation{Wallenberg Centre for Quantum Technology, Department of Microtechnology and Nanoscience, Chalmers University of Technology, Sweden , SE-412 96 G\"{o}teborg, Sweden}

\begin{abstract}
Quantifying non-Gaussianity is essential for assessing the computational power of bosonic quantum systems. We introduce a framework to witness and quantify non-Gaussianity from displaced parity measurements, and we directly apply it to experimental data obtained with superconducting cavities.
Our method estimates fidelities with target pure states from the displaced parity measurements, enabling certification of non-Gaussianity via the stellar rank. We further introduce refined witnesses computed by measuring the value of the Wigner function at a few points, providing efficient certification tools outperforming existing ones. Applying these techniques to Fock, binomial, cat, and GKP experimental data, we rigorously certify non-Gaussian resources and establish a stellar rank of $8$ for a GKP state, setting a new benchmark across all quantum platforms. Our methods enable a rigorous quantification of non-Gaussian quantum resources directly applicable to superconducting devices.
\end{abstract}

\maketitle

% ============================================================
\section{Introduction}
\label{sec:introduction}
% ============================================================

Bosonic systems have emerged as promising candidates for large-scale quantum computing \cite{Weedbrook:12, pfister2019continuous, fukui_building_2022-1}. These include superconducting platforms, which currently offer state-of-the-art error-correction capabilities \cite{ofek2016extending, reglade2024quantum, putterman2025hardware, brock2025quantum}, photonic platforms, which enable deterministic generation of entanglement at unprecedented scales \cite{yoshikawa2016invited, aghaee2025scaling}, as well as trapped-ion \cite{Fluhmann, deNeeveHome} and optomechanical \cite{barzanjeh2022optomechanics} platforms.

In bosonic quantum computing architectures, non-Gaussianity ---namely, the property of the bosonic field of exhibiting fluctuations that are not described by Gaussian statistics--- is necessary in order to achieve exponential quantum computational advantage \cite{Bartlett2002,Mari2012, veitch2014resource,chabaud2023resources}, despite not being sufficient \cite{garcia2020efficient,hahn2025bridging,xmtw-g54f}. Therefore, being able to assess non-Gaussianity in concrete experimental platforms is crucial in order to ensure their potential usefulness for computation. 

Several resource measures have been developed to quantify non-Gaussianity in bosonic systems \cite{albarelli2018resource,chabaud2020stellar,Hahn2025classicalsimulation,Hahn2026assessingnon,hahn2025measuringnongaussianitycorrelation,mele2025symplectic}. Among these, the stellar rank \cite{chabaud2020stellar} provides a non-Gaussian resource monotone with operational significance for quantum computation, in that it quantifies the increase in computational cost for classical simulation \cite{chabaud2023resources}. It is therefore important to provide methods for its certification and quantification. However, current approaches \cite{chabaud2020certification} do not exploit the specific capabilities of platforms in which the Wigner function can be directly measured, such as superconducting platforms.

Here, we introduce two protocols to witness the stellar rank that are applicable to any platform that allows direct access to the Wigner function at arbitrary phase-space points. The first protocol is based on fidelity estimation, which relies on sampling the phase-space using a target quasi-probability distribution, in this case the Wigner function of the target state~\cite{fawzi2024optimal}, i.e. the state that we wanted to generate in an experiment. This protocol is fairly general and straightforward to apply ---once a reasonable target state is identified--- however its sample complexity scales with the negativity of the Wigner function of the target state. The second protocol is based on optimized witnesses that only require measuring the Wigner function at properly chosen very few phase-space points. The advantage of this protocol is  that it only requires a low number of measurements, whereas its drawback is that the witness strongly depends on the target state, and its theoretical identification is computationally demanding and "ad-hoc" for each target state. However, we also show that, for two-point witnesses, we can establish analytically a general connection between the potential of the witness of detecting stellar rank, and the range of the Wigner function, i.e. the difference between its maximum and minimum point.
We prove that the two protocols are of immediate practical relevance by simulating their application based on data coming from real experiments \cite{Kudra}. In particular, our simulations provide strong theoretical and numerical evidence of a stellar rank of 8.  This value goes beyond the value of 6 that was previously detected in Ref.\cite{rahman2025genuine}, in terms of genuine quantum non-Gaussianity, the two notions being equivalent for pure states \cite{lachman2019faithful, asenbeck2025hierarchical}.
Finally, for rotationally symmetric states, we introduce a phase-space averaged few-point witness which, as the name suggests, averages over phase space information.  As a result, this witness yields lower thresholds, as compared to the non phase-averaged version, thereby enabling the certification of comparatively higher stellar rank for rotationally symmetric states.

The rest of the paper is structured as follows: after providing the necessary background in \cref{sec:preliminaries}, we present our fidelity-based approach in \cref{sec:fidelity_wit} and demonstrate it with experimental data; then, in \cref{sec:displaced_parity_wit}, we introduce few-points witnesses, allowing us to witness the stellar rank of experimental states by checking only a few values of their Wigner function. We also present a detailed analysis of the two-points witness. In \cref{sec:phase_averaged_witness}, we introduce the phase-averaged witness and highlight its advantages for rotationally symmetric states. Finally, we conclude in \cref{sec:conclusion} with a discussion and present open research directions.

% ============================================================
\section{Preliminaries on continuous-variable systems}
\label{sec:preliminaries}
% ===================================================
We start by presenting an introduction of the theoretical framework underlying our analysis. We first establish the notation and fundamental concepts underpinning continuous-variable systems. Subsequently, we introduce the stellar representation of quantum states and its usage for quantifying non-Gaussianity via the stellar rank. Finally, we present the class of quantum states considered in this work and outline their role as benchmarks for the proposed techniques.

\subsection{General CV background}
We consider an infinite dimensional Hilbert space $\mathcal{H}$ with the associated creation and annihilation operators $\hat{a}^\dagger, \hat{a}$ fulfilling the commutation relation
\begin{align}
    [\hat{a},\hat a^\dag]=\hat{\mathbb 1}.
\end{align}
We denote by $\bm{\hat a}=(\hat a_1,\dots,\hat a_m)^T$ the tuple of annihilation operators over $m$ modes.

Gaussian unitaries are defined as
\begin{align}
    \hat U_G&=e^{-i \hat H_G},
\end{align}
where $\hat H_G$  is an operator at most quadratic in the creation and annihilation operators. Such evolutions transform the creation and annihilation operators linearly.

For a general Gaussian operator the Hamilton operator $\hat H_G$ can be written as
\begin{align}
    \hat{H}_G=\bm{\alpha}^T\bm{\hat{a}^\dagger}+\bm{\hat{a}^\dagger}U \bm{\hat{a}}+\bm{\hat{a}^\dagger} S \bm{\hat{a}^\dagger}+h.c.,
\end{align}
for $\bm{\alpha}\in\mathds{C}^m$ and where $U,S$ are $m\times m$ complex matrices for $m$ modes.
Any Gaussian unitary can be decomposed using the Euler decomposition as
\begin{align}
    \hat U_G= \hat{U} \hat{S}(\bm{\xi}) \hat{D}(\bm{\alpha})   \hat{V},
\end{align}
where $\hat U, \hat V$ are passive Gaussian unitaries, $\hat{S}(\bm{\xi})=\bigotimes_{i=1}^m \hat{S}(\xi_i)$ single mode squeezing and $\hat{D}(\bm{\alpha})=\bigotimes_{i=1}^m \hat{D}(\alpha_i)$ displacement operations.
Passive Gaussian unitaries are generated by Hamiltonians of the form $\hat{H}_p=\bm{\hat{a}^\dagger}U\bm{\hat{a}} +h.c.$ and conserve the total number of excitations.

As a consequence any generic single-mode Gaussian unitary can be decomposed into a sequence of displacements, squeezing and phase shifts:
\begin{align}
    \hat{D}(\alpha)&=e^{\alpha\hat{a}^\dag -\alpha^* \hat{a}},\\
    \hat{S}(\xi) &= e^{\frac{1}{2}(\xi \hat{a}^{\dag2}-\xi^* \hat{a}^2)},\\
    \hat{R}(\phi)&=e^{i \phi\,\hat{a}^\dagger \hat{a}}.
\end{align}
For the multi-mode case we only require the addition of beam splitter unitaries
\begin{align}
    \hat{U}_{BS}(\theta)=e^{\theta(\hat{a}^\dagger_1\hat{a}_2-\hat{a}_1\hat{a}^\dagger_2)},
\end{align}
where the variable $\theta$ sets the transmissivity of the beam splitter $\tau = \cos(\theta)^2$, yielding a balanced beam-splitter if $\tau = \frac{1}{2}$ or equivalently $\theta = \frac{\pi}{4}$.
As a consequence, a generic $m-$mode unitary can be decomposed using a sequence of beam splitter unitaries and single mode phase shifters, squeezing and displacement. 

Gaussian states are defined as 
\begin{align}
    \rho_G=\frac{e^{-\beta \hat H_G}}{\Tr[e^{-\beta\hat H_G}]},
\end{align}
which reduces for $\beta\rightarrow\infty$ to pure states. 
We can write pure Gaussian states as 
\begin{align}
    \ket{\psi_G}&=\hat U_G \ket{0^{\otimes m}}
    = \hat{U} \hat{S}(\bm{\xi}) \hat{D}(\bm{\alpha})\ket{0^{\otimes m}}.
\end{align}

Alternatively, continuous-variable quantum systems can be described in phase space via quasi-probability distributions. Among these, the Wigner function plays a particularly central role, as it provides a complete representation of the quantum state while closely resembling a classical probability distribution. For an $m$-mode state $\rho$, it is defined as
\begin{align}
W_{\rho}(\bm{\alpha}) = \frac{2^m}{\pi^m}\Tr\left[\hat{D}(\bm{\alpha})\hat \Pi\hat{D}^\dagger(\bm{\alpha}) \rho\right],
\end{align}
where $\hat{D}(\bm{\alpha})$ is the multi-mode displacement operator and $\hat \Pi=\left(\sum_{n=0}^\infty (-1)^n \dyad{n}\right)^{\otimes m}$ denotes the multi-mode parity operator.
The Wigner function admits many features of a classical phase-space distribution, such as normalization and marginal distributions yielding actual probability-density functions. However, in contrast to classical distributions, it may take on negative values, which is a direct sign of non-classicality. The degree of Wigner negativity is therefore directly linked to fundamental quantum resources and has been widely used as an indicator of non-Gaussianity.
Beyond its conceptual significance, the Wigner function also offers an experimentally accessible characterization tool. In many state-of-the-art platforms, most notably circuit QED architectures, the natural coupling between qubits and cavity modes enables a direct measurement of the Wigner function \cite{PhysRevLett.78.2547}. In such schemes, displacements of the cavity mode combined with qubit readout allow one to reconstruct the phase-space distribution point by point, thereby providing a powerful method for quantum state tomography \cite{doi:10.1126/science.1243289}.

\subsection{Stellar rank}
Alternatively, pure quantum states can be represented by holomorphic functions, known as stellar functions. This framework offers a complementary viewpoint to phase-space quasi-probability distributions, while being especially well-suited for analyzing non-Gaussian features of pure quantum states. 
For any pure quantum state $\ket{\psi}=\sum_{\bm{n}\geq0} \psi_{\bm{n}}\ket{n}$, where $\bm{n} = (n_1,n_2,\dots,n_m) \in \mathbb{N}^m$, we can associate a stellar function~\cite{bargmann,segal}, defined as
\begin{equation}
    F^\star_{\bm\psi}(\bm z):=\sum_{\bm n\ge\bm 0}\frac{\psi_{\bm n}}{\sqrt{n_1!\dots n_m!}}\prod_{i=1}^m  z_i^{n_i},
\end{equation}
for all $\bm z = (z_1,\dots,z_m) \in\mathbb C^m$.
This representation is inherently connected with the Husimi-Q function as 
\begin{align}
    Q(\bm{\alpha})_\psi=\frac{\e^{-\abs{\bm{\alpha}}^2}}{\pi^m}\abs{F^\star_{\bm\psi}(\bm \alpha)}^2.
\end{align}
As long as the decomposition $\ket{\psi}=\sum_{\bm{n}\geq0} \psi_{\bm{n}}\ket{n}$ is finite, the stellar function decomposes as $P\times G$, where $P$ is a multivariate polynomial of total degree $r^\star$ and $G$ is a multivariate Gaussian function.
Thus up to a Gaussian pre-factor the stellar function is determined by its zeroes.
The collection of these zeros, called the stellar constellation, encodes the essential non-Gaussian features of the state.
This perspective leads naturally to the notion of \textit{stellar rank}, defined as the total degree $r^\star$ of the polynomial $P$ obtained after factoring out the Gaussian contribution $G$ from the stellar function~\cite{chabaud2020stellar}.
The stellar rank serves as an operational measure of non-Gaussianity. Aside from invariance under Gaussian unitary operations and faithfulness, the stellar rank is even non-increasing under post-selected Gaussian protocols~\cite{Hahn2026assessingnongaussianquantumstate}.

Additionally, the stellar rank also admits a strong operational interpretation. The stellar function provides a direct way of expressing the quantum state in terms of creation operators:
$\ket{\bm\psi}=F^\star_{\bm\psi}(\hat a^\dag_1,\dots,\hat a^\dag_m)\ket{\bm0}$. 
This expression highlights that a state of stellar rank $n$ necessarily requires the action of at least $n$ creation operators, supplemented by Gaussian unitary operations, in order to be prepared.

The stellar rank can, in general, be unbounded; for instance any cat state possesses infinite stellar rank due to the non-polynomial structure of their stellar functions. Moreover, the stellar rank is not robust in the following sense. For example, the family of states $\ket{\psi}=(1-\delta)\ket{0}+\delta\ket{n}$ has stellar rank $n$ $\forall \delta>0$.
In practice, however, for $\delta$ sufficiently close to zero, no experiment would be able to reliably distinguish $\ket{\psi}$ from the vacuum state. Thus, while mathematically exact, the stellar rank can overestimate the effective non-Gaussianity present in experimentally relevant regimes.

This observation motivates the introduction of the $\epsilon$-approximate stellar rank, which provides a more fine-grained measure of non-Gaussianity~\cite{Hahn2026assessingnongaussianquantumstate}.
For $0\le\epsilon\le1$, the $\epsilon$-approximate stellar rank of an $m$-mode density operator $\rho$ is defined as
\begin{equation}\label{eq:def-asr}
    r^\star_\epsilon(\bm\rho):=\inf_{\bm\tau,F(\bm\rho,\bm\tau)\ge1-\epsilon}r^\star(\bm\tau),
\end{equation}
where $F(\bm\rho,\bm\tau)=\mathrm{Tr}(\sqrt{\sqrt{\bm\rho}\bm\tau\sqrt{\bm\rho}})^2$ denotes the fidelity. The $\epsilon$-approximate stellar rank can be computed numerically using a connection to a hierarchy of stellar fidelities~\cite{Hahn2026assessingnongaussianquantumstate}
\begin{equation}\label{eq:stellarF}
    f_n^\star(\bm\rho):=\sup_{\bm\tau,r^\star(\bm\tau)\le n}F(\bm\rho,\bm\tau).
\end{equation}
The stellar fidelities and the $\epsilon$-approximate stellar rank are then related via
\begin{equation}
    f^\star_n(\bm\rho)\ge1-\epsilon\Leftrightarrow\forall\epsilon'>\epsilon,\,r^\star_{\epsilon'}(\bm\rho)\le n.
\end{equation}

Several techniques have been introduced in the literature to experimentally witness the stellar rank. In Ref.~\cite{Fiurasek:22}, a general framework for constructing stellar-rank witnesses was introduced, along with two explicit witness operators tailored to cat states and Fock states. This approach was subsequently refined in Ref.~\cite{provazník2026witnessesnongaussianfeatureslower}, where additional non-linear terms were incorporated. While these extensions can enhance detection capabilities, they also lead to more challenging optimization problems.
In general, the practical implementation of such witnesses can be demanding, as their feasibility depends strongly on the operations available in a given experimental setup. In Ref.~\cite{wassner2025detectingquantumnongaussianitysingle}, an alternative method was proposed that relies on homodyne measurements, which are readily available in optical setups. This approach enables the estimation of the zeros of the wavefunction, which can in turn be used to witness the stellar rank. Finally, \cite{lee2026efficient} and \cite{PhysRevLett.114.190402} propose methods to witness the non-Gaussianity of states by probing their Wigner function, but these witnesses do not tell us anything about the amount of non-Gaussianity.

In contrast to the methods discussed above, we propose stellar rank witnessing techniques based on displaced parity measurements, which are naturally suited to superconducting architectures. In the following, we introduce two distinct approaches for constructing witness operators: one based on fidelity estimation, and another that relies directly on a small number of phase-space points measurements. In particular, the latter method is obtained by specializing the general framework introduced in Ref.~\cite{Fiurasek:22} to witnesses based on displaced parity measurements.

\subsection{States of interest}
\label{sec:states}
To benchmark our methods against experimentally relevant scenarios, we first  focus on the family of quantum states realized in Ref.~\cite{PRXQuantum.3.030301}.
In that work, the authors generated genuine Wigner negative states in a 3D microwave cavity.
Specifically, aside from Fock states, they realized a binomial state
\begin{align}
\label{eq:binom_state}
    \ket{\beta}=\frac{1}{\sqrt{2}}\qty(\ket{0}+\ket{4}),
\end{align}
an odd-parity cat state
\begin{align}
\label{eq:cat_state}
     \ket{\alpha_0^-}=\frac{1}{\sqrt{\mathcal{N}}}\left(\ket{\alpha_0=\sqrt{2}}-\ket{\alpha_0=-\sqrt{2}}\right)
\end{align}
and a approximate Gottesman–Kitaev–Preskill (GKP) zero codeword state~\cite[Eq. 7.7a]{Albert2018}
\begin{align}
\label{eq:GKP_state}
    \ket{0}_{\text{GKP}}=\frac{1}{\sqrt{\mathcal{N'}}}\sum_{n=-\infty}^\infty \e^{-\frac{\pi}{2} (2n\Delta)^2}  D\qty(\sqrt{\frac{\pi}{2}}2n) S(-\log\Delta)  \ket{0}
\end{align}
with $\Delta=0.35$.
All of these states play a central role in CV quantum computing. Binomial states and Schrödinger cat states are prominent examples of rotationally symmetric bosonic codes, which enable the encoding of logical qudits into CV systems \cite{grimsmo_quantum_2020}. GKP states, while belonging to a distinct family of bosonic codes, also provide a qudit encoding, albeit based on grid-like phase-space structures \cite{gottesman_encoding_2001}. Crucially, each of these states encode discrete quantum information into continuous variables and thereby serves as a building block for error-corrected quantum computation.

% ============================================================
\section{Fidelity-based witnesses}
\label{sec:fidelity_wit}
% ============================================================
In this section, we introduce a first method to experimentally witness the amount of non-Gaussianity a state contains, which we refer to as \textit{fidelity-based witnesses}. 

We consider the following setting: an experimentalist aims to prepare a target pure non-Gaussian state $\ket{\psi}$. Due to inevitable experimental imperfections, however, the actual prepared state is a mixed state $\rho$, which is not known exactly. The central question is then: to what extent does the realized state $\rho$ exhibit non-Gaussianity?

\begin{table*}
    \centering
    \begin{tabular}{c|c c c c c}
         State & $F$ with $n=10^4$  & $F$ with $n=10^5$ & $F$ with $n=10^6$ & $\epsilon$ for  $\delta=0.1$ & $\epsilon$ for $\delta=0.01$ \\\hline
        Binomial state &  0.9340 & 0.9359 &0.9372  & 0.0077 & 0.0109\\
         Cat state & 0.9477 & 0.9510  & 0.9503 & 0.0064& 0.0091 \\
         GKP state & 0.9443 & 0.9483  &  0.9497& 0.0080 & 0.0113\\
         Fock $\ket{1}$ & - & -  &  0.9591& 0.0061 & 0.0087\\
         Fock $\ket{2}$  & -& -  &  0.9957& 0.0074 & 0.0105
    \end{tabular}
    \caption{Estimated fidelity of the experimental data from \cite{PRXQuantum.3.030301} with the targeted states. Errors are computed with maximal samples. Reported experimental fidelities are 0.94, 0.96, 0.94, respectively~\cite{PRXQuantum.3.030301}.}
    \label{tab:exp_fidelity}
\end{table*}

\begin{figure*}
    \centering
    \includegraphics[width=1\textwidth]{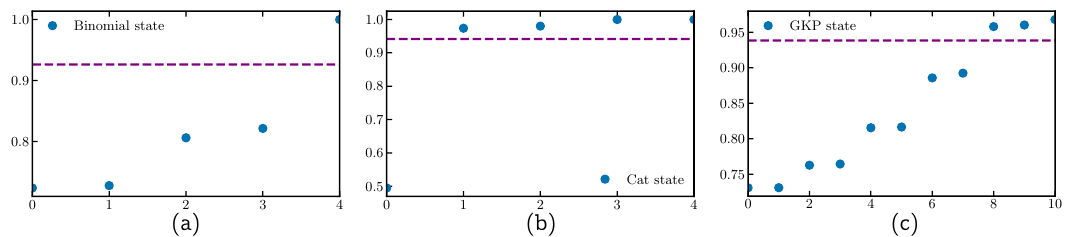}
    \caption{Stellar fidelity profile for the binomial state of Eq.~\eqref{eq:binom_state}(a),  the cat state of Eq.~\eqref{eq:cat_state}(b) and the  GKP state of Eq.~\eqref{eq:GKP_state}(c). The horizontal line is the fidelity estimate minus half the error for a failure probability of $1\%$. The point where the horizontal line intersects with the stellar fidelity profile is the certified stellar rank rounded up.}
    \label{fig:fid_wit}
\end{figure*}

\subsection{General approach}
Our first approach establishes a fidelity-based certification framework that directly links experimentally accessible overlap estimates to the non-Gaussian structure of the target state. The key idea is to combine fidelity estimation with the stellar fidelity profile of $\ket{\psi}$~\cite{Hahn2026assessingnongaussianquantumstate}, which captures the fine-grained non-Gaussian features of the state. This connection enables us to convert lower bounds on the fidelity into rigorous and quantitative guarantees on the stellar rank.
Concretely, we compute the stellar fidelity profile of $\ket{\psi}$ and combine it with a black-box fidelity estimation protocol~\cite{fawzi2024optimal} (see also App.~\ref{ap:fidelity_est}). This yields a practically implementable certification method that avoids full state tomography while still providing strong, resource-theoretic statements about the prepared state.

The basic idea of the protocol is to use the Wigner function of state $\ket{\psi}$ and sample coordinates $\bm{\alpha}_i$ based on that Wigner function using the probability distribution 
\begin{align}
    p(\bm\alpha)=\frac{\abs{W_{\dyad{\psi}}(\bm\alpha)}}{\norm{W_{\dyad{\psi}}}_1},
\end{align}
where $\norm{W_{\dyad{\psi}}}_1=\int \dd\bm{\alpha}\abs{W_{\dyad{\psi}}(\bm{\alpha})}$is the Wigner negativity.
For a chosen $\bm\alpha$ we experimentally measure the Wigner function of state $\rho$ at point $\bm\alpha$, namely $W_{\rho}(\bm\alpha)$.
We can then obtain one sample by computing
\begin{align}
    X=W_{\rho}(\bm\alpha) \norm{W_{\dyad{\psi}}}_1 \text{sign}(W_{\dyad{\psi}}(\bm\alpha)).
\end{align}
The expectation value of $X$ is then exactly the fidelity between state $\ket{\psi}$ and $\rho$
\begin{align}
    \mathds{E}(X)=\pi^m\int \dd\bm{\alpha} W_{\dyad{\psi}}(\bm \alpha) W_\rho(\bm \alpha)=F(\dyad{\psi},\rho).
\end{align}
Using this procedure we can upper bound the error $\epsilon$ of the fidelity with
\begin{align}
    \epsilon\leq \norm{W_{\dyad{\psi}}}_1 \qty(2^3  N^{-1} \log(\frac{1}{\delta}))^{\frac{1}{2}},
\end{align}
where $N$ is the number of samples, and $\delta$ the failure probability.
To allow for a fair comparison between the stellar fidelity and the estimated fidelity, we subtract the error committed in the estimation, $\epsilon$, from the obtained fidelity $F'(\dyad{\psi},\rho)=F(\dyad{\psi},\rho)-\epsilon$ with probability at least $1-\delta$.
We then compare the fidelity $F'(\dyad{\psi},\rho)$ with the stellar fidelity profile. The stellar rank we witness this way is the smallest $n$ such that $F'(\dyad{\psi},\rho) \leq f^\star_n(\psi)$. We  stress that the stellar fidelities are obtained through optimisation over a non-convex space and so we cannot be sure in general that the numerical value obtained corresponds to the global optimum. However, for many practical examples, especially in the case of single modes, the optimisation space is small enough so that we can have high confidence.

A key advantage of this approach is that it circumvents the need for full quantum state tomography, which is experimentally demanding and scales unfavorably with system size and energy~\cite{mele2024learningquantumstatescontinuous}. Instead, by relying only on carefully sampled values of the Wigner function, the method provides certification of non-Gaussianity where the sample complexity --- and therefore in a sense the experimental feasibility --- increases with the 1-norm of the Wigner and yields thus better scaling than tomography.

Note that this procedure for estimating the stellar rank works even in the case where the experimentally generated state is radically different than the target. In that case, though, the samples from the experimentally generated Wigner function will give a lower estimated fidelity, and hence a lower certified stellar rank. The fidelity-based protocol is thus more robust to experimental inaccuracies in the state preparation. The number of samples to obtain the stellar rank scales, however, with the Wigner negativity of the target state.

\subsection{Application to experimental data}
\label{sec:fidelity_exp_data}

We now apply the fidelity-based witnesses to experimental data corresponding to the states generated in Ref.~\cite{PRXQuantum.3.030301}. To this end, we numerically simulate the experimental protocol outlined in the previous section. Specifically, we first compute the stellar fidelity profile of the states introduced in Section~\ref{sec:states}. Next, we evaluate the Wigner function of the target pure states and select a set of sampling points ${\alpha_i}$ in phase space. The experimental data is provided as a reconstructed density matrix, from which we compute the Wigner function at the phase-space points ${\alpha_i}$.
We use the experimentally obtained data instead of simulated data in order to use states with genuine experimental noise and test our protocols against real noise.
Finally, we use these sampled values to estimate the fidelity between the experimentally realized state and the target state.
In Table~\ref{tab:exp_fidelity} we present the fidelity obtained using the experimental data and the black box fidelity estimation protocol.

Using the estimated fidelities and the stellar fidelity profiles we can now estimate the stellar rank of the experimentally generated states.
As a start we compare the estimated fidelity of Fock states $\ket{1}$ and $\ket{2}$ with the stellar fidelities. The stellar fidelities for $\ket{1}$ are 
$0.48,1$ and for $\ket{2}$ are $0.38,0.56,1$, where the numbers are the fidelities for a general rank $0,1,2$ state respectively.
By comparing these numbers with the estimated fidelity while subtracting half of the estimated error, we observe that  the the estimated fidelity for $\ket{1}$ is larger than $0.48$ and we can therefore obtain s a stellar rank 1, but not stellar rank $2$ as the estimated fidelity would need to be $1$. For $\ket{2}$ the estimated fidelity is larger than $0.56$. Thus we obtain a stellar rank of $2$, but not $3$.
For Fock states, our approach certifies the correct stellar rank showing the high quality of the experimentally generated states as well.

Furthermore, we investigate states that are widely used for CV error correction.
The first state of this family we consider is the binomial state defined in Eq.~\eqref{eq:binom_state}.
The stellar fidelity profile, together with the estimated fidelity, for the binomial state can be found in Figure~\ref{fig:fid_wit}(a).
We see that the estimated fidelity minus half of the error is larger than the stellar fidelity of rank $3$, but below $4$. We obtain therefore a stellar rank of $4$. As we can see from Eq.~\eqref{eq:binom_state} in theory a stellar rank of $4$ is possible. Thus we obtain the true stellar rank even using experimental data, proving again the high quality of the experimentally prepared states.

The case of cat states is a bit more involved.
The stellar rank of any cat state is infinite, independent of the magnitude of the state. However, by looking at stellar fidelity profiles, larger magnitudes lead to a slower convergence of the fidelity to $1$~\cite{Hahn2026assessingnongaussianquantumstate} making it experimentally feasible to detect higher stellar ranks.
In Figure~\ref{fig:fid_wit}(b) we show the stellar fidelity profile with the estimated fidelity for the cat state~\eqref{eq:cat_state}. We observe that the estimated fidelity lies between the stellar fidelity of a rank 0 and a rank 1 state. We obtain thus a stellar rank of at least $1$.

The last state we are considering is the GKP state defined in Eq.~\eqref{eq:GKP_state}.
We provide the stellar fidelity profile with the estimated fidelity for the GKP state in Figure~\ref{fig:fid_wit}(c). We see that the estimated fidelity is between the stellar fidelity corresponding to rank $7$ and rank $8$. Therefore we obtain a stellar rank of at least 8. This means that we need a superposition of Fock states including Fock state $\ket{8}$ to obtain a state that has the same fidelity with the GKP state defined in Eq.~\eqref{eq:GKP_state}.

As we have seen in this section, using stellar fidelity profiles and black box fidelity estimation one can robustly estimate the stellar rank of experimental states.
The only requirement is that one has knowledge of what state is targeted in the experiment.
For Fock states and binomial states, the protocol can certify the theoretical maximum.
The stellar fidelities for the cat state are too close to $1$ making it challenging to use those to estimate the stellar rank beyond revealing the non-Gaussian nature of the generated state. For the GKP state we obtain a stellar rank of $8$.

%============================================================
\section{Few-points witnesses}
\label{sec:displaced_parity_wit}
% ============================================================

The method presented in the previous section based on direct fidelity estimation is more efficient than full state tomography. However, the required number of samples can still be very high if the desired state has a large amount of Wigner negativity. Moreover, direct fidelity estimation requires probing the Wigner function at a wide range of randomly sampled points $\bm\alpha$.
In this section we introduce another family of witnesses for the stellar rank, which we call \textit{few-points witnesses}. By construction, these stellar rank witnesses require measuring the Wigner function at only a handful of points, and are thus tailored to displaced parity measurements. This approach is similar to \cite{lee2026efficient} and \cite{PhysRevLett.114.190402} where the authors witness non-Gaussianity by probing few points of the Wigner function, but their approach does not witness the amount of non-Gaussianity.

\subsection{Definition}

\noindent Formally, few-points witness operators are defined as
\begin{align}
   \hat W(\bm a,\bm\alpha)=\sum_{j=1}^Na_j \hat D(\alpha_j)\hat \Pi \hat D^\dag(\alpha_j),
\end{align}
where $\hat \Pi$ is the parity operator, $a_j \in \mathbb{R}$ and $\sum_{j=1}^N \abs{a_j}=1$.

To measure such a witness operator, one simply measures the Wigner function at the points $\alpha_j$ and adds the estimates weighted by $a_j$. For a single point, $O(\frac{1}{\epsilon^2})$ samples are sufficient to obtain an estimate up to additive error $\epsilon$ by Hoeffding's inequality, which amounts to a total of $\frac{N}{\epsilon^2}$ measurements for an $N$-point witness.

The threshold to certify a stellar rank greater than $k$ using these witnesses is given by:
\begin{align}\label{eq:optimization_problem}
    w_k(\bm a,\bm\alpha)=\sup_{r^\star(\rho)\le k}\mathrm{Tr}[\hat W(\bm a,\bm\alpha)\rho].
\end{align}
Note that by this definition, the witnesses are are invariant under affine symplectic transformations.
Following~\cite{Fiurasek:22} this can be equivalently expressed as 
\begin{align}
    w_k(\bm a,\bm\alpha)=\sup_{\hat G\in\mathcal G}[\mathrm{maxeig}(\hat\Pi_k \hat G \hat W(\bm a,\bm\alpha) \hat G^\dag\hat\Pi_k)];
\end{align}
%involving the maximum eigenvalue of the operator $\hat\Pi_k \hat G \hat W(\bm a,\bm\alpha) \hat G^\dag\hat\Pi_k)$.
here $\hat\Pi_k=\sum_{l=0}^k\ket k\!\bra k$ is the projector on the space of states with at most $k$ photons, $\hat\Pi_k \hat G \hat W(\bm a,\bm\alpha) \hat G^\dag\hat\Pi_k$ is a $(k+1)\times(k+1)$ matrix, $\mathrm{maxeig}$ denotes the largest eigenvalue, and the optimisation is over Gaussian unitaries. We can simplify this expression further to allow for numerical computation: we show in~\cref{ap:proofII} and~\cref{ap:compact} that
\begin{widetext}
    \begin{align}
    \label{eq:threshold}
     w_k(\bm a,\bm\alpha)=\sup_{x,y\in\mathbb R,r\ge0,\phi\in [0,2\pi)]}\left[\mathrm{maxeig}\left(\sum_{j=1}^Na_j\sum_{0\le m,n\le k}\pi_{m,n}(x+iy+\alpha_j\cosh r+\alpha_j^* \e^{\ii \phi}\sinh r)|m\rangle\langle n|\right)\right],
\end{align}
where
    \begin{align}
    \label{eq:pi_c}
\pi_{m,n}(z)
=
 e^{-2|z|^2}
 \sum_{p=0}^{\min(m,n)}
 \frac{\sqrt{m!n!}\,(-1)^p}
 {p!(m-p)!(n-p)!}
 (2z)^{m-p}(2z^*)^{n-p}.
\end{align}
Note that $\pi_{m,n}(z)$ are the Laguerre 2D-functions presented in~\cite{Wunsche1998}.
\end{widetext}

Similar to stellar fidelity witnesses \cite{PRXQuantum.2.020333,Hahn2026assessingnongaussianquantumstate}, a limitation of this technique is that the numerical bounds derived do not come with a certificate. Since the optimisation is non-convex, we cannot be sure in general that the numerical value obtained corresponds to the global optimum. For many practical examples, however,
the optimisation space is small enough so that we can have high confidence about the numerical bounds obtained. This is the case in particular for most single-mode examples and witnesses for low stellar ranks.

On the other hand, the family of few-points witnesses $\hat W(\bm a,\bm\alpha)$ allows for considerable of freedom in choosing: (i) the number of points $N$, (ii) the location of the points $\bm\alpha$, and (iii) the weights $\bm a$. We address the question of how to choose these parameters in the next subsection.

\subsection{Choosing the witness' parameters}
\label{subsec:choose}

In this subsection we outline our strategy for choosing a few-points witness' parameters.

Firstly, we note that if the weights of a few-points witness are non-negative, it is necessary to pick $N>1$. Indeed, for a single point $\alpha$, the coherent state $\ket\alpha$ saturates the expectation value of the witness $\hat W(1,\alpha)$. Moreover, for a single-point with negative weights, the witnesses reduces to a test of Wigner negativity. Hereafter, we consider witnesses based on two or more points. We provide a more detailed analysis of  two-point witnesses in subsection \ref{sec:two_point_witness}. 
The main advantage of few-points witnesses is that they require very few measurements of the Wigner function to witness the stellar rank.
As a result, this makes measuring these witnesses much less experimentally demanding compared to the fidelity-based method.
On the other hand, these witnesses must still be tailored to a target state. 
Indeed, few-points witnesses, even if they may have a non-trivial threshold value $w_k(\bm a,\bm\alpha)$, will in general not successfully witness the stellar rank of a state, unless the structure of the state is taken into account. 

Similar to fidelity-based witnesses, we can use the information about the structure of a target state rather than that of the actual state in the experiment. Then, a general strategy is to associate positive weights to a few local maxima of the Wigner function of the target state and negative weights to a few local minima. In particular, we pick positive weights if $W(\alpha_i)\geq0$ and negative weights if $W(\alpha_i)<0$ (as we show in subsection \ref{sec:two_point_witness}, this is actually necessary to do for two-point witnesses). Finally, although we fix the magnitude of all weights to be equal for simplicity, we note that it may be beneficial to numerically optimize the weights in practice.

In the next subsection, we apply such tailored few-points witnesses to experimental data and do numerical analysis to witness stellar rank. Following that, in subsection \ref{sec:two_point_witness}, we do a more extensive analysis of the two-point witness which highlights analytically the relation between the witness value and the nature of the Wigner function. Finally, in \cref{app:robustness}, we analyse the robustness of these few-points witnesses both to noise and imprecision in the location of the points. In both cases, we find that the few-points witnesses exhibit some level of robustness, which decreases as the features of the target states sharpen.  
%\oliver{Would it makes sense to slightly extend the robustness discussion in the main text, i.e. what we learned etc? That part is a few pages in the appendix but small in the main text.}

\subsection{Application to experimental data}
\label{sec:parity_exp_data}

In this subsection, we introduce few-points witnesses tailored to the target states described in \cref{sec:states}, according to the strategy outlined in the previous subsection \ref{subsec:choose}: at least two points, chosen at local maxima with positive weight and local minima with negative weight of equal magnitudes. Importantly, in all cases we pick the location of the points using only the ideal target state, rather than based on the experimental state.
Then, we apply the few-points witnesses to experimental data from~\cite{PRXQuantum.3.030301}. Once again, the experimental data is provided as a reconstructed density matrix, from which we compute the expectation values of the witnesses in classical post-processing.

\begin{figure*}[t]
\centering
\begin{minipage}{0.33\textwidth}
    \centering
    \includegraphics[width=\linewidth]{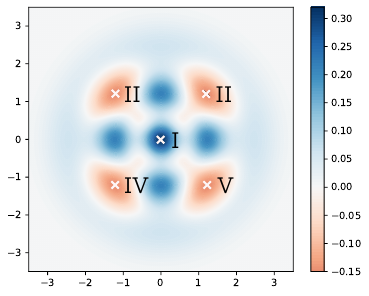}
    \caption{Wigner function of the studied binomial state. The crosses with roman numbers indicate the location of points of the Wigner functions used to define the corresponding few-points witnesses.}
    \label{fig:bin_state}
\end{minipage}
\hfill
\begin{minipage}{0.65\textwidth}
    \centering
    \begin{tabular}{c|c c c c | c c}
         $k$& 0 &1 &2 &3 & Th. &Exp.\\\hline
        Points: I,II &0.5& 0.7795& 0.8771& 0.9209 &0.7330 & 0.6583 \\
         Points: I,II,III& 0.3333& 0.5083  &  0.7685 &  0.7930 & 0.6441& 0.5730\\
         Points: I,II,III,IV,V & 0.2&0.3999&0.4611&0.4795 &0.5729 &0.5200
    \end{tabular}
    \captionof{table}{Table of thresholds values $w_k$ for the binomial state in \cref{eq:binom_state} and few-points witness expectation values. Each line corresponds to a few-points witness chosen by picking a subset of points from Figure~\ref{fig:bin_state} indicated in the first column. An expectation value (Th.\ or Exp.) higher than a threshold value $w_k$ certifies a stellar rank $>k$. Here, stellar rank $\ge1$ (resp.\ $\ge1$) is certified via the first witness for the ideal (resp.\ experimental) state, stellar rank $\ge2$ (resp.\ $\ge2$)  while the maximal stellar rank $4$ is certified via the last witness for both states.}
    \label{tab:thresh_values_bin}
\end{minipage}
\end{figure*}

\subsubsection{Binomial state}

The points of the Wigner function we use to build the few-points witness for the binomial state from \cref{eq:binom_state} are shown in \cref{fig:bin_state}. The corresponding threshold values and the values of the witness for the theoretical binomial state and the experimental binomial state are reported in \cref{tab:thresh_values_bin}.

In this case, only two points are necessary to witness non-zero stellar rank. In particular, we pick the maximal value (the origin) and the smallest value of one of the negative regions, as can be seen in \cref{fig:bin_state}, when picking point I and II.
As we can see in \cref{tab:thresh_values_bin}, using this configuration the witness certifies a stellar rank $\ge1$ for the theoretical state and for the experimental state.
The target state has stellar rank $4$, and if we pick points (I,II,III,IV and V) we can detect the maximum possible stellar rank of $4$ both for the ideal binomial state and the experimental state. Thus, by only measuring three points of the Wigner functions, we can witness the stellar rank of $4$ of a binomial state.

\subsubsection{Cat state}

\begin{figure*}[t]
\centering
\begin{minipage}{0.33\textwidth}
    \centering
    \includegraphics[width=\linewidth]{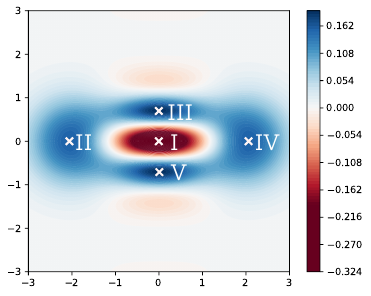}
    \caption{Wigner function of the studied cat state. The crosses with roman numbers indicate the location of points of the Wigner functions used to define the corresponding few-points witnesses.}
    \label{fig:cat_state}
\end{minipage}
\hfill
\begin{minipage}{0.65\textwidth}
    \centering
\begin{tabular}{c| c c c c  | c c}
        $k$ & 0 &1 &2 &3  & Th. &Exp.\\\hline
        
            Points: I,II &0.5&0.7795&0.8771&0.9209&0.7464&0.7026\\
            Points: I,III &0.5&0.7795&0.8771&0.9209&0.7997&0.7337\\
             Points: I,III,V &0.6667&0.7693&0.903&0.9239&0.7329&0.6801\\
         Points: I,II,III,IV& 0.5&0.6023&0.6651& 0.7603&0.6463&0.6098\\
        Points: I,II,III,IV,V &0.4&0.5563&0.6104&0.6948&0.6369&0.5989\\
   
    \end{tabular}
    \captionof{table}{Table of thresholds values $w_k$ for the cat state in \cref{eq:cat_state} and few-points witness expectation values. Each line corresponds to a few-points witness chosen by picking a subset of points from Figure~\ref{fig:cat_state} indicated in the first column. An expectation value (Th.\ or Exp.) higher than a threshold value $w_k$ certifies a stellar rank $>k$. For instance, stellar rank $\ge3$ (resp.\ $\ge2$) is certified via the witness in the last line for the ideal (resp.\ experimental) state.}
    \label{tab:thresh_values_cat}
\end{minipage}
\end{figure*}

The points of the Wigner function we use to build the few-points witness for the cat state from \cref{eq:cat_state} are shown in \cref{fig:cat_state}(b). The corresponding threshold values and the values of the witness for the theoretical cat state and the experimental cat state are reported in \cref{tab:thresh_values_cat}.

We find that one can witness a non-trivial stellar rank for cat states using only two points of the Wigner function if we choose points I and II or I and III (see \cref{fig:cat_state}).
As we can see in \cref{tab:thresh_values_cat}, the witness can detect stellar rank 1 for both the experimental and theoretical cat state for I and II, but stellar rank 2 for the theoretical state in the case of points I and III.

Using more points improves the capabilities of the witness: using four different points (I, II, III and IV), we can witness stellar rank 2 for both the theoretical and experimental sate; using five different points (I, II, III, IV and V) we can witness stellar rank 3 for the theoretical state but still a stellar rank 2 for the experimental state. However, we note that simply increasing the number of points does not necessarily improve the performance of the witness. %For instance, if we choose the points I and III or I, III and V, we only witness stellar rank $1$ for the experimental state, while the witness using less points can detect stellar rank 2 for the theoretical state.

Note that in this case, few-points witnesses allow for detecting a higher stellar rank as compared to the fidelity-based witness. Strikingly, with the latter, even after sampling one random  million points, the witnessed stellar rank for the case of the experimentally generated cat state was equal to one, while here a handful of carefully chosen samples of the same Wigner function make it possible to obtain a higher stellar rank. The advantage of the witness-based method is due to the careful choice of the points to be measured, while for the fidelity-based methods the used fidelities are close to 1 even for low stellar ranks making it hard to surpass them experimentally.

\subsubsection{GKP state}
\begin{figure*}[t]
\centering
\begin{minipage}{0.33\textwidth}
    \centering
    \includegraphics[width=\linewidth]{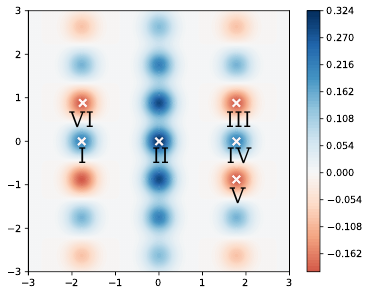}
    \caption{Wigner function of the studied GKP state. The crosses with roman numbers indicate the location of points of the Wigner functions used to define the corresponding few-points witnesses.}
    \label{fig:gkp_state}
\end{minipage}
\hfill
\begin{minipage}{0.65\textwidth}
    \centering
\begin{tabular}{c|c c c c c c c| c c}
        $k$ & 0 &1 &2 &3 &4 &5 &6 &Th. &Exp. \\\hline
        Points: II,III,V &0.3333&   0.6664&0.6667& 0.6776& 0.7331& 0.7487&-&0.7353&0.6493\\
        Points: II,V,VI& 0.3333&0.5081&0.7685&-&-&-&-&0.7353&0.6686\\ 
         Points: I,II,III,IV,V & 0.6&0.6&0.6&0.6&0.6855 &0.68845 &0.7327& 0.7119&0.6240\\
    \end{tabular}
    \captionof{table}{Table of thresholds values $w_k$ for the GKP state in \cref{eq:GKP_state} and few-points witness expectation values. Each line corresponds to a few-points witness chosen by picking a subset of points from \cref{fig:gkp_state} indicated in the first column. An expectation value (Th.\ or Exp.) higher than a threshold value $w_k$ certifies a stellar rank $>k$. The first witness can potentially witness a stellar rank of $\geq5$, but only manages to witness stellar rank $\geq 1$ for the experimental state. The second witness certifies a stellar rank of $\geq2$ for both. The last witness certifies a stellar rank of $\geq 6$ (resp. $\geq 4$).}
    \label{tab:thresh_values_gkp}
\end{minipage}
\end{figure*}

The points of the Wigner function we use to build the few-points witness for the GKP state from \cref{eq:GKP_state} are shown in \cref{fig:gkp_state}. The corresponding threshold values and the values of the witness for the theoretical GKP state and the experimental GKP state are reported in \cref{tab:thresh_values_gkp}.

As can be seen in \cref{tab:thresh_values_gkp}, three points (II, III and V) are potentially sufficient to witness a stellar rank of $5$ for the theoretical state and a stellar rank of $1$ for the experimental state. If we instead choose the points II,V,VI then we can witness a stellar rank of $2$ for both, but loose the ability to certify a high stellar rank for the theoretical state
Adding two more points such that we now have (I,II,III,IV and V) improves the witnessing capabilities, as can be seen in \cref{tab:thresh_values_gkp}. 
We are then able to witness a stellar rank fo 6 and 4 respectively.
In particular, finding the optimal sampling points is non-trivial, and an interesting problem stemming from our work is determining how does the optimal number of sampling points and their location depend on the state which stellar rank we wish to certify.

\subsection{Two-point witness}\label{sec:two_point_witness}
In this subsection, we present a further investigation of few-point witnesses constructed from exactly two-points and see for which states this approach is particularly advantageous.  
The appeal of a few-point witness lies in the fact that only a small number of phase-space points need to be measured to witness stellar rank of the measured state. As mentioned previously, a single-point cannot be used as a witness. So the  minimal extension beyond a single-point is a witness based on two points, which will now focus on.

One interesting feature to note is that any two-point witness can be transformed into any other two-point witness with the same weights using only Gaussian operations. This follows from the fact that we can induce a general symplectic transformation on the displacement variable using rotations and squeezing operations. If we consider a nonzero displacement operator $\hat D(\beta)$, with $\beta=|\beta|e^{i\phi}\neq0$, we can transform it into some other nonzero displacement operator $\hat D(\alpha)$, where $\alpha=|\alpha|e^{i\theta}\neq0$, by conjugation with a Gaussian unitary $\hat{U}_S=\hat{R}(\theta)\hat{S}^\dagger(r)\hat{R}^\dagger(\phi)$ with $r=\ln(|\beta|/|\alpha|)$:
\begin{equation}
    \hat{U}_S\hat{D}(\beta)\hat{U}_S^\dagger = \hat{D}(\alpha). 
\end{equation}
Since $\hat U_S$ is composed only of rotations and squeezing operations, it preserves the parity operator, i.e., $\hat{U}_S\hat{\Pi}\hat{U}_S^\dagger=\hat{\Pi}$. This gives that any nonzero displacement can be transformed into any other while leaving parity invariant. Since symplectic transformations are invertible, no finite Gaussian unitary can map a nonzero displacement to the origin. Thus, the origin can only be approached asymptotically in the infinite-squeezing limit $r\to\infty$.

Using this, any two-point witness can be related to any other two-point witness defined with the same weights by Gaussian conjugation, which can be seen by the following calculation, for some two-point witness $\hat{W}((a_1,a_2),(\alpha_1,\alpha_2))$:
\begin{widetext}
\begin{align}
    \hat{W}((a_1,a_2),(\alpha_1,\alpha_2)) & = a_1\hat{D}(\alpha_1) \hat{\Pi} \hat{D}^\dagger(\alpha_1)+a_2\hat{D}(\alpha_2) \hat{\Pi} \hat{D}^\dagger(\alpha_2)\\
    & = \hat{D}(\alpha_1)\left(a_1 \hat{\Pi} +a_2\hat{D}(\alpha_2-\alpha_1) \hat{\Pi} \hat{D}^\dagger(\alpha_2-\alpha_1)\right)\hat{D}^\dagger(\alpha_1) \\
    & = \hat{D}(\alpha_1)\hat{U}_S^\dagger\left(a_1 \hat{\Pi} +a_2\hat{D}(\alpha_2'-\alpha_1') \hat{\Pi} \hat{D}^\dagger(\alpha_2'-\alpha_1')\right)\hat{U}_S\hat{D}^\dagger(\alpha_1) \\
    & = \hat{D}(\alpha_1)\hat{U}_S^\dagger\hat{D}^\dagger(\alpha_1')\left(a_1 \hat{D}(\alpha_1')\hat{\Pi}\hat{D}^\dagger(\alpha_1') +a_2\hat{D}(\alpha_2') \hat{\Pi} \hat{D}^\dagger(\alpha_2')\right)\hat{D}(\alpha_1')\hat{U}_S\hat{D}^\dagger(\alpha_1) \\
    & = \hat{G}\hat{W}((a_1,a_2),(\alpha_1',\alpha_2'))\hat{G}^\dagger,
\end{align}
\end{widetext}
where $\hat{D}(\alpha_2-\alpha_1)=\hat{U}_S\hat{D}(\alpha_2'-\alpha_1')\hat{U}_S^\dagger$, and we have defined the overall Gaussian  operator $ \hat{G} = \hat{D}(\alpha_1)\hat{U}_S^\dagger\hat{D}^\dagger(\alpha_1')$.
\begin{table*}[t]
\centering
\begin{tabular}{c|*{11}{D}}
$k$ & 0 & 1 & 2 & 3 & 4 & 5 & 6 & 7 & 8 & 9 & 10 \\
\hline
$w_k(\bm a,\bm\alpha)$ & 0.5  & 0.7795 & 0.8771 & 0.9209 & 0.9443 & 0.9583 & 0.9674 & 0.9736 & 0.9781 & 0.9815 & 0.9841 \\
\end{tabular}
\caption{Thresholds for the frist 11 stellar ranks corresponding to the two-point witness with weights $a_1=1/2$ and $a_2=-1/2$. }
\label{tab:two-point thresholds}
\end{table*}

An important consequence of this is that the threshold of a two-point witness only depends  on the choice of weights. Since the thresholds are obtained through an optimization over all Gaussian unitaries, Gaussian-equivalent witnesses must share the same thresholds. As shown above, any two-point witnesses with the same weights are Gaussian-equivalent. Consequently, the thresholds need only be computed once for a given choice of weights, independent of the phase-space locations at which the witness is evaluated. 

Additionally, the coalescing point limit $\delta=\alpha_2-\alpha_1\to0$ yields further insight. If we define a witness operator $\hat{W}((a_1,a_2),(0,\delta))=a_1\hat{\Pi}+a_2\hat{D}(\delta)\hat{\Pi}\hat{D}^\dagger(\delta)$ --- which, by the previous argument, is Gaussian-equivalent to any other two-point witness with the same weights --- we see that 
\begin{align}
    & \lim_{\delta\to0}\hat{D}(\alpha)\hat{U}_S\hat{W}((a_1,a_2),(0,\delta))\hat{U}_S^\dagger\hat{D}^\dagger(\alpha)\\
    & =(a_1+a_2)\hat{D}(\alpha)\hat{U}_S\hat{\Pi}\hat{U}_S^\dagger\hat{D}^\dagger(\alpha)=(a_1+a_2)\hat{W}(1,\alpha),\label{eq:lim two-p wintess}
\end{align}
where we have used $\hat{U}_S\hat{\Pi}\hat{U}_S^\dagger=\hat{\Pi}$. Thus, in this limit, every two-point witness reduces to a single-point witness with a proportionality constant.

When $a_1,a_2\ge0$, an important consequence of the limit in Eq.~\eqref{eq:lim two-p wintess} is that the two-point witness is Gaussian equivalent to a single-point witness in the limit of coincident-point. Therefore, the two-point witness inherits the single-point thresholds in this limit. Thus, a two-point witnesses with non-negative weights cannot be used to certify stellar rank.

In light of these observations, we restrict our attention throughout the remainder of this subsection to equal-magnitude weights with opposite signs, namely $a_1=\tfrac{1}{2}$ and $a_2=-\tfrac{1}{2}$, which coincides with the convention adopted in earlier analysis. In Table~\ref{tab:two-point thresholds}, we list the first $11$ thresholds corresponding to these choice of weights. As we have seen in subsection~\ref{sec:parity_exp_data}, two point are enough to witness non-trivial stellar rank for Fock state $\ket{1}$ and the considered odd-parity cat state.

Since the thresholds are independent of the phase-space points used to construct the witness, it is natural to ask whether there is a general criterion that classifies the maximum expectation value for a given state, and thereby directly determines the stellar rank of the state. For equal-magnitude weights, we find that the maximum expectation value for a state $\rho$ is proportional to the range of its Wigner function, as demonstrated below:
\begin{widetext}
\begin{align}\label{eq:two_point_range}
    \max_{\alpha_1,\alpha_2\in\mathbb{C}}\Tr\!\left[\hat{W}((1/2,-1/2),(\alpha_1,\alpha_2))\rho\right] & = \frac{1}{2}\max_{\alpha_1\in\mathbb{C}}\left(\Tr\left[\hat{D}(\alpha_1)\hat \Pi\hat{D}^\dagger(\alpha_1) \rho\right]\right) + \frac{1}{2}\max_{\alpha_2\in\mathbb{C}}\left(-\Tr\left[\hat{D}(\alpha_2)\hat \Pi\hat{D}^\dagger(\alpha_2) \rho\right]\right) \\ 
    & = \frac{\pi}{4}\left(\max_{\alpha_1\in\mathbb{C}}W_{\rho}(\alpha_1)-\min_{\alpha_2\in\mathbb{C}} W_{\rho}(\alpha_2)\right) \\
    & = \frac{\pi}{4}\mathcal{R}(W_{\rho}),
\end{align}
\end{widetext}
where the range of the Wigner function is defined as $\mathcal{R}(W_{\rho})=\max_{\alpha\in\mathbb{C}} W_{\rho}(\alpha)-\min_{\alpha\in\mathbb{C}} W_{\rho}(\alpha)$.

This result provides a simple interpretation of the two-point witness: its optimal expectation value is determined entirely range of the Wigner function. Since the thresholds in Table~\ref{tab:two-point thresholds} depend only on the choice of weights and not on the specific phase-space points, the stellar rank that can be certified is determined directly by the Wigner function structure. We already observed such a behavior for the best performing two-point witness for Fock and cat state in subsec.~\ref{sec:parity_exp_data}.
Consequently, states with a larger Wigner-function range are generally better suited for certifying higher stellar rank using the two-point witness.

In Ref.~\cite{Chabaud_2024}, it was shown that the Wigner function satisfies the bound $-2/\pi \leq W_{\rho}(\alpha) \leq 2/\pi$. For states with a non-negative Wigner function, this reduces to $0 \leq W_{\rho}(\alpha) \leq 2/\pi$, implying that the Wigner function range satisfies $\mathcal{R}(W_{\rho}) \leq 2/\pi$. Since the maximum value of the two-point witness is given by $(\pi/4)\mathcal{R}(W_{\rho})$, it follows that the witness value cannot exceed $1/2$. This coincides with the $k=0$ threshold in Table~\ref{tab:two-point thresholds}. Consequently, the two-point witness cannot certify nontrivial stellar rank for states whose Wigner function is non-negative. On the other hand, observing a nontrivial stellar rank with the equal weight two-point witness implies the presence of Wigner negativity. The two-point witness can therefore be viewed as an indirect test for the presence of Wigner negativity.

\subsubsection{Fock states and odd-parity cat states}
We will now investigate how the two-point witnesses perform for ideal Fock states and ideal odd-parity cat states beyond
the ones studied above. The choice of considering here ideas states is dicated by teh fact that  we do not  have access to experiential data for Fock states above 2 and cat states with a general amplitude. The Wigner functions of the Fock state $\ket{n}$ and the odd-parity cat state $\ket{\alpha_0^-}$ from Eq.~\eqref{eq:cat_state} are respectively given as
\begin{equation}
    W_{\ket{n}}(\alpha) = \frac{2}{\pi}(-1)^ne^{-2|\alpha|^2}L_n(4|\alpha|^2),
\end{equation}
and 
\begin{align}
    W_{\ket{\alpha_0^-}}(\alpha)& =\frac{4}{\pi \mathcal{N}}e^{-2(\alpha_R^2+\alpha_I^2)} \notag\\
    & \times\left(e^{-2\alpha_0^2}\cosh(4\alpha_0\alpha_R)-\cos(4\alpha_0\alpha_I)\right),
\end{align}
where $\alpha=\alpha_R+i\alpha_I$. The Fock state Wigner function is obtained immediately by using the displacement in Fock basis from Eq.~\eqref{eq:disp Fock basis}. The derivation of the cat state Wigner function is presented in Appendix~\ref{app:odd even Wigner}. 
\begin{figure*}[t]
    \centering
    \includegraphics[width=1.0\linewidth]{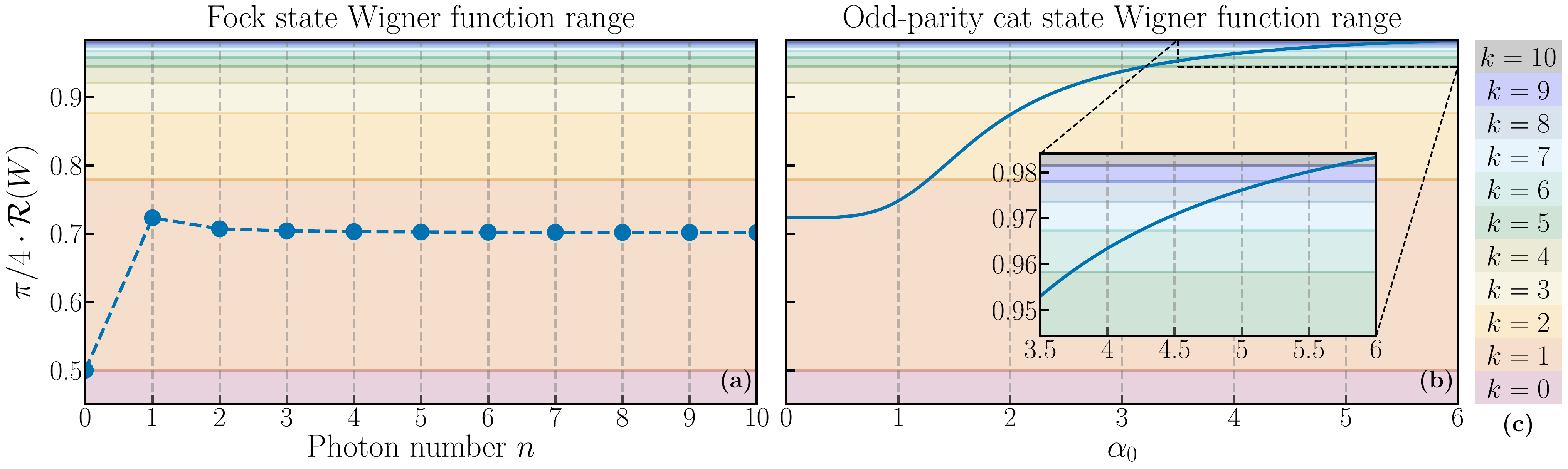}
    \caption{(a) Range of the ideal Fock state Wigner function for photon number $0\leq n\leq10$. (b) Range of the ideal odd-parity cat as a function of the amplitude $\alpha_0$. (c) The color used in the backgrounds of the (a) and (b) plots to represent the threshold values from Table~\ref{tab:two-point thresholds}.} 
    \label{fig:Fock odd cat ranges}
\end{figure*}
In Figure~\ref{fig:Fock odd cat ranges}, we plot the range of Fock states and odd-parity cat state and compare them with the thresholds. 

For all non-vacuum Fock states in Figure~\ref{fig:Fock odd cat ranges}(a), the two-point witness is able to certify a stellar rank of at most $1$ for all photon numbers. This behavior can be understood from the corresponding Wigner-function ranges, which attain their maximum for the single-photon state and then decrease monotonically with increasing photon number. As a result, the best achievable expectation value of the with two-point witness with equal magnitude weights also decreases with photon number and never exceeds the threshold required to certify a higher stellar rank. Based on the observed trend, we expect this behavior to persist for larger photon numbers. 

Figure~\ref{fig:Fock odd cat ranges}(b) shows the range of the Wigner function for odd-parity cat states as a function of the amplitude $\alpha_0$. As $\alpha_0$ increases, the Wigner-function range grows monotonically. Consequently, the maximum expectation value attainable with the two-point witness also increases, allowing the certification of progressively higher stellar ranks. In particular, for $\alpha_0 = 5$, the witness certifies a stellar rank of $7$, surpassing the highest stellar rank certified in the previous sections. Increasing the amplitude further to $\alpha_0 = 6$ enables in principle the certification of stellar rank $9$.

In Appendix~\ref{app:robustness}, we further investigate the performance of the two-point witness by analyzing its ability to certify the stellar rank of odd- and even-parity cat states in the presence of loss and displacement errors in the measurement.
%----------------------------------------------------------
\section{Phase-averaged witness}\label{sec:phase_averaged_witness}
From the previous section, it is clear that to certify a stellar rank that is larger than one for Fock states, we need to construct witnesses with more than two points. In section \ref{appsec:fock7_few_points} of the appendix, we demonstrate how to witness stellar rank 7 for Fock state $\ket{7}$ using 22 points. 

For the construction of a well-performing witness, we observe that increasingly more points are necessary to certify the full stellar rank of rotationally symmetric Fock states. In addition to that, the choice of the point phases has an influence on the thresholds of the witness even given the rotational symmetry of the target states. This demonstrates that the geometric arrangement of the phase-space points constitutes a non-trivial optimization problem, and substantial improvements can be achieved even without changing the points' absolute values in phase-space where the minima and maxima of the Fock states are located.

We therefore introduce a further, phase-averaged, witness. The motivation for defining a phase-space averaged witness follows from the observation that, for a rotationally symmetric state, phase-space points with the same radius would result in the same measurement value of the Wigner function. Treating such points as independent sampling points increases the witness complexity with little gain of information. This observation suggests constructing the witness itself in a rotationally symmetric way. For a sampling radius $r_1$, instead of evaluating the displaced parity operator at one phase-space point, we average over all rotations of that point,
\begin{equation}\label{eq:phase_averaged_single}
    \hat{W}_{PA}(a_1,r_1)
    =
    \int_0^{2\pi}\frac{d\theta}{2\pi}
    a_1
    \hat{D}(r_1e^{i\theta})
    \hat{\Pi}
    \hat{D}^\dagger(r_1e^{i\theta}).
\end{equation}
A phase-averaged few-point witness can then be constructed by combining several radii with different weights,
\begin{equation}
    \hat{W}_{PA}(\boldsymbol{a}, \boldsymbol{r})
    =
    \sum_{j=1}^{N} \hat{W}_{PA}(a_j, r_j).
\end{equation}

As proved in section \ref{appsec:phase_averaged_witness} of the appendix, the structure of phase-averaged witness allows the analytical structure of the original threshold optimization to be retained. The threshold for a certification level $k$ can again be written as an optimization over Gaussian transformations followed by the largest eigenvalue of the corresponding truncated operator. The key difference is that rotational symmetry removes one degree of freedom from the Gaussian optimization. In the phase-averaged case, the rotation angle of the Gaussian transformation does not need to be optimized. The rotational symmetry of the witness allows this parameter to be removed, leaving only the squeezing magnitude and the complex displacement as independent optimization parameters:
\begin{eqnarray}
    w_k^{PA}(\boldsymbol{a}, \boldsymbol{r})
    &=& \nonumber \\
    && \hspace{-20mm}\sup_{r_s\in[0,\infty),\,\gamma\in\mathbb{C}}
    \lambda_{\max}
    \left[
        \hat{\Pi}_k
        \hat{D}(\gamma)\hat{S}(r_s)
        \hat{W}_{PA}(\boldsymbol{a}, \boldsymbol{r})
        \hat{S}^\dagger(r_s)\hat{D}^\dagger(\gamma)
        \hat{\Pi}_k
    \right]. \nonumber\\
    \label{eq:pa_optimization}
\end{eqnarray}
Thus, the four real parameters of the general Gaussian transformation are reduced to three, that is, the optimization is performed over a smaller parameter space, which reduces the computational cost and allows more independent optimization runs to be performed within the same computational time. This is relevant because the optimization problem is non-convex and the numerical thresholds do not come with a strict guarantee that the global optimum has been found. Repeating the optimization from different initial conditions therefore provides an important consistency check. Reducing the dimensionality of the optimization makes it possible to increase the number of such independent attempts and thereby obtain greater confidence in the numerical thresholds.

The phase-averaged construction also changes the interpretation of the sampling procedure. In the original few-point witness, each selected phase-space point corresponds to a specific displacement. For a rotationally symmetric state, however, points related by a rotation are equivalent. The phase-averaged witness associates a measurement with a radius and combines the information from all angular positions at that radius. This makes the construction naturally adapted to states such as Fock states, whose Wigner functions are themselves rotationally symmetric.

There is, however, a trade-off to this construction. Phase-averaging removes information about the angular structure of the Wigner function. As a result, a phase-averaged witness is not expected to perform equally well for states whose non-Gaussian features strongly depend on the phase-space angle. In addition, phase-averaging can reduce the amount of information available for distinguishing states with different stellar ranks. Therefore, the witness adaptation does not automatically imply a higher certified stellar rank. 

The phase-averaged witness should therefore be understood as a symmetry-adapted alternative rather than a general improvement of the original few-point construction. Its main analytical benefit is the reduction of the optimization problem, while its suitability for a given state depends on whether the information discarded by phase-averaging is relevant for stellar-rank certification. For these reasons, we use the phase-averaged witness primarily as a tailored construction for rotationally symmetric states rather than as a general replacement for the few-point witness.

\begin{figure*}[t]
\centering
\begin{minipage}[t]{0.23\textwidth}
    \vspace{0pt}
    \centering
    \includegraphics[width=\linewidth]{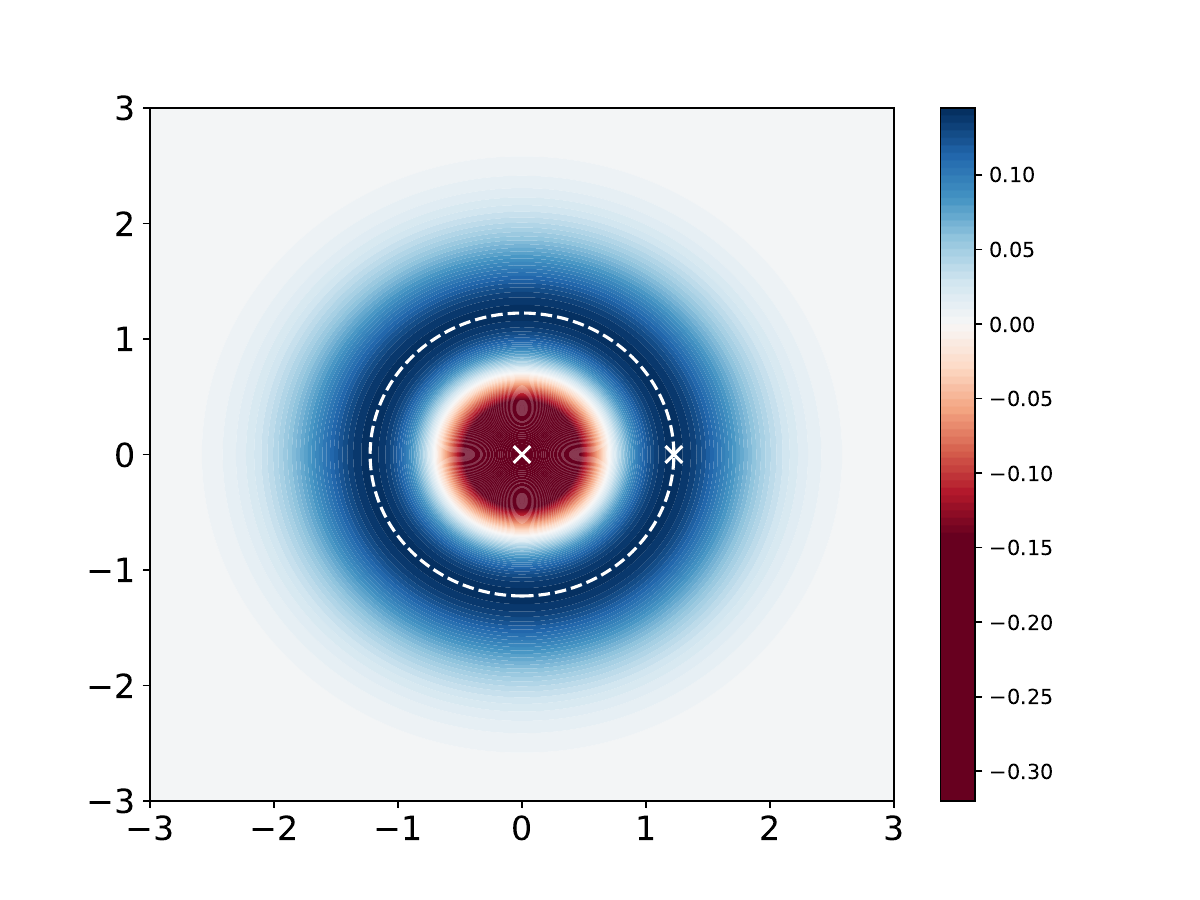}
    \caption{Wigner function for the Fock state $n=1$,
    with the radii and points used to construct the
    phase-averaged witness.}
    \label{fig:fock1_pa}
\end{minipage}
\hfill
\begin{minipage}[t]{0.23\textwidth}
    \vspace{0pt}
    \centering
    \includegraphics[width=\linewidth]{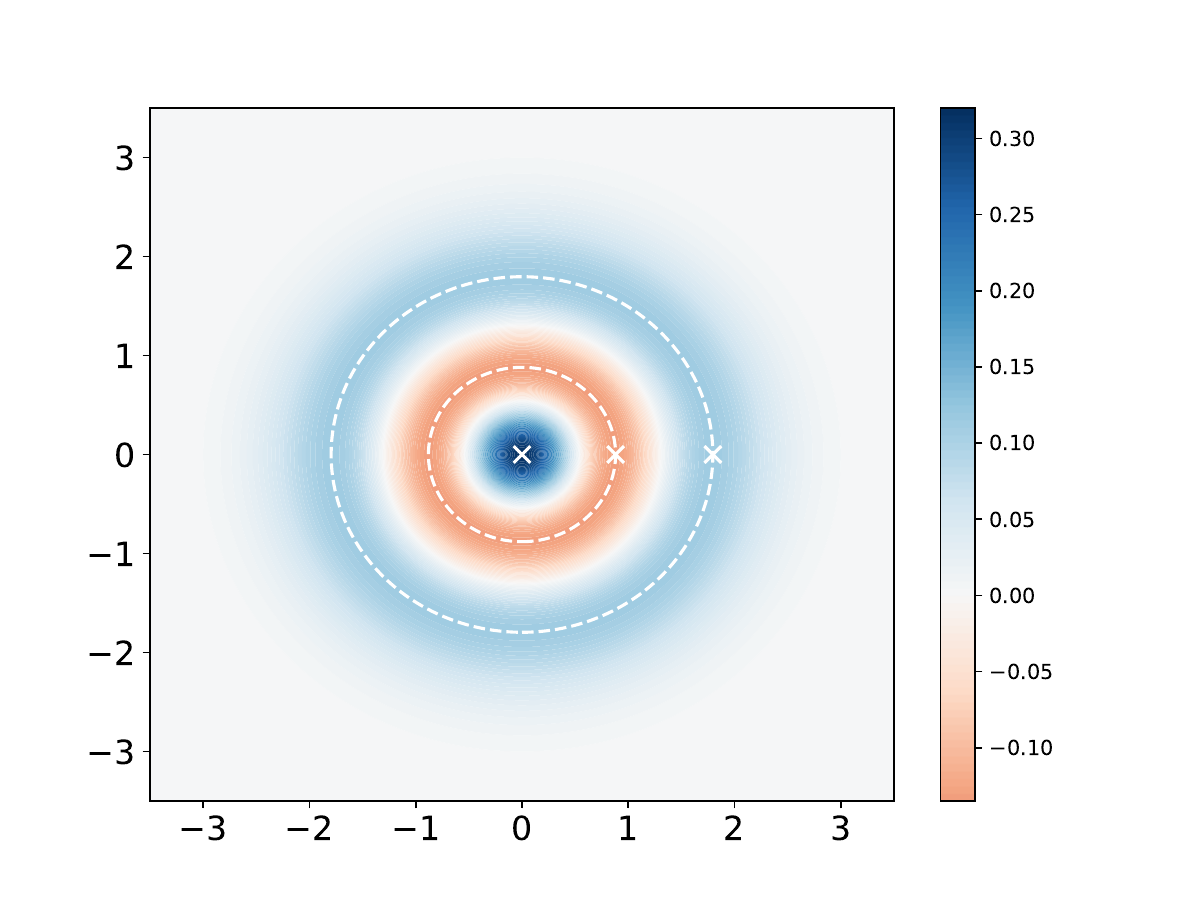}
    \caption{Wigner function for the Fock state $n=2$,
    with the radii and points used to construct the
    phase-averaged witness.}
    \label{fig:fock2_pa}
\end{minipage}
\hfill
\begin{minipage}[t]{0.50\textwidth}
    \vspace{0pt}
    \centering
    \begin{tabular}{c|c|c|c|c|c}
        $n$ & $w^{PA}_0$ & $w^{PA}_1$ & $w^{PA}_2$
            & Th. & Exp. \\ \hline
        1 & 0.1377 & 0.7231 & --
          & 0.7231 & 0.6799 \\ \hline
        2 & 0.3333 & 0.3333 & 0.5893
          & 0.5893 & 0.5628 \\ \hline
    \end{tabular}
    \captionof{table}{Witness thresholds and theoretical (Th.)
    and experimental (Exp.) expectation values for the
    phase-averaged witness applied to Fock states $n=1$ and $n=2$.
    Here, Th.\ and Exp.\ denote
    $\mathrm{Tr}(\hat{W}_{PA}(\boldsymbol{a},\boldsymbol{r})\rho)$
    and
    $\mathrm{Tr}(\hat{W}_{PA}(\boldsymbol{a},\boldsymbol{r})\rho_{exp})$,
    respectively, where $\rho=\ket{n}\bra{n}$ is the ideal
    Fock state and $\rho_{exp}$ is the experimentally
    reconstructed state from~\cite{Kudra}.
    An expectation value higher than $w^{PA}_k$
    certifies a stellar rank $r^*> k$.}
    \label{tab:thresholds_pa_exp}
\end{minipage}
\end{figure*}

As an illustration, here we present our results on witnessing stellar rank using experimental data of Fock states $\ket{1}$ and $\ket{2}$ provided in \cite{PRXQuantum.3.030301} using the phase-averaged witness. The corresponding threshold values and witness expectation values are shown in Table~\ref{tab:thresholds_pa_exp}. For Fock state $\ket{1}$, the phase-averaged witness has a threshold of $0.7231$ for $k=1$, while the experimental witness expectation values are $0.6799$ for the original reconstructed state and $0.6926$ after phase averaging. Since both values exceed the $k=0$ threshold of $0.1377$, stellar rank 1 can be certified. Similarly, for Fock state $\ket{2}$, the phase-averaged witness has a $k=1$ threshold of $0.3333$, while the experimental expectation values are $0.5628$ and $0.5863$ for the original and phase-averaged states, respectively. Therefore, both certify a stellar rank of at least 2 and using a phase-averaged witness, we can certify the actual stellar rank of both $\ket{1}$ and $\ket2$.

Going beyond the regime of the experimentally available data that we have at our disposal, in sec.~\ref{appsec:fock7_phase_averaged} of the Appendix we demonstrate how the phase-averaged witness can be used to certify a stellar rank of seven for the Fock state $\ket{7}$ and compare its performance with that of the few-point witness. We also explain there how Eq.~(\ref{eq:phase_averaged_single}) can be computed numerically on a given Wigner function.

Finally we note that, in practice, the phase averaged witness could be probed either by discarding the phase reference when performing the measurements of the displaced parity, or by sending the state through a dephasing channel before performing the measurements of the displaced parity. 
%-------------------------------------------------------
\section{Conclusion and Outlook}
\label{sec:conclusion}
% ============================================================

In this work, we have introduced three new experimentally-friendly methods to witness stellar rank, a measure of non-Gaussianity yielding a computational resource. The methods are based on measuring the displaced-parity, and are hence particularly amenable to experimental platforms where the Wigner function can be directly sampled, such as superconducting platforms. The first method is based on fidelity estimation, and typically requires a few thousands to a million samples. The second protocol is based on optimized witnesses that only require measuring  the Wigner function at few properly chosen phase-space points, making it particularly practical from an experimental point of view; in particular, for states with high Wigner function range,  witnesses based on sampling the Wigner function in just two-points have the potential of certifying high stellar rank (e.g., 9 for a Fock state of amplitude $\alpha =6$).  Finally, our third protocol is particularly well-suited for rotationally symmetric states and is computed by averaging the value of the Wigner function over all phase space points at a particular radius. Testing our first method on experimental data for reconstructed Wigner functions allowed us to provide strong theoretical and numerical evidence that the experiment in Ref.~\cite{PRXQuantum.3.030301} achieved a stellar rank of 8. Since stellar rank is an important and platform-independent quantifier of computational resources \cite{chabaud2023resources}, our methods offer a way to facilitate platform comparisons by providing a means to estimate the stellar rank in systems where the Wigner function can be directly measured at arbitrary phase-space points. 

Our work opens several research directions. First, the scalability of resource certification methods remains a pressing issue: our current approach witnesses single-mode non-Gaussianity and it would be interesting to generalize it to witness multi-mode non-Gaussianity \cite{mele2025symplectic}.
Additionally, analyzing the robustness of phase-averaged witness to photon loss and displacement error, similar to how we did for few-points witness, also represents a promising research direction. More generally, while non-Gaussianity constitutes a fundamental resource for universal bosonic quantum computing, it is by no means the only one: extending our methodology to other resources, such as entanglement, is essential to establish a comprehensive, platform-agnostic witnesses toolbox for benchmarking bosonic quantum computers. Availability of such a toolbox is critical to assess progress and compare platforms' capabilities for quantum computing, guiding  their technological development.

\section*{Acknowledgments}
We thank the experimental group of Simone Gasparinetti and Axel Eriksson, at Chalmers, and the other authors of Ref.\ \cite{PRXQuantum.3.030301}, for having made available the experimental data to us. 
 U.C., V.U., A.F.\ and G.F.\ acknowledge funding from the European Union’s Horizon Europe Framework Programme (EIC Pathfinder Challenge project Veriqub) under Grant Agreement No.\ 101114899.
G.F.\ and A.M.\ acknowledge financial support from the Swedish Research Council (Vetenskapsrådet) through the project grant DAIQUIRI, as well as from the Olle Engkvist foundation. G.F.\ and S. R.-K.\ acknowledge support from the Knut and Alice Wallenberg Foundation through the Wallenberg Center for Quantum Technology (WACQT). O.H.\ acknowledges the support from CREST Grant Number JPMJCR23I3, Japan.

\section*{Authors contributions}
O.H., A.M., and V.U.\ carried out the majority of the analytical and numerical calculations; %O.H. did the analytical derivation of the original few-point witness, as well as of the fidelity-based witness; V.P. and A.M.  ran numerical simulations of the few-point witnesses; M.E. A.M developed the formalism for the two-point witness; V.P. analyzed witness robustness and finalized the writing of the manuscript;
M.E.\ advanced the analytical and numerical treatment of the few-point witness, developed the analytical and numerical analysis of the two-point and phase-averaged witnesses, and performed part of the robustness analysis reported in the Appendix; U.C., A.F., G.F., and S.R.-K.\ supervised the research; U.C.\ conceived and initiated the project, including the original fidelity-based and few-point witness approaches; S.R.-K.\ conceived the phase-averaged and two-point witnesses; A.F.\ helped shape the presentation of the work and its relevance to experimental implementations; G.F.\ coordinated the international collaboration and oversaw the project workflow; all authors discussed the results and contributed to the writing and revision of the manuscript.

\section*{AI disclosure}
 All analytical calculations were performed manually, except the derivation of Eq.~\eqref{eq:pi_c} in Appendix \ref{ap:compact}, where AI was used to find a compact expression after a more involved expression was derived manually in appendix \ref{ap:proofII}; Claude Code has been used to optimise the numerical codes used to derive the results.

%%%%%%%%%%%%%%%%%%%%%%%%%%%%%%%%%%%%%%%%%%%%%%%%%%%%%%%%%%%%%%%%%%%%%%%%%%%%%%%%%%%%
\newpage
\appendix
\widetext

\section{Blackbox fidelity estimation}
\label{ap:fidelity_est}
In this section we will give a short introduction to the blackbox fidelity estimation protocol introduced in Ref.~\cite{fawzi2024optimal}.
Afterwards, we explain how we implemented it in practice to sample from experimental data.
\subsection{Algorithm}
We want to estimate the fidelity between a known state $\dyad{\psi}$ and an unknown state $\sigma$
\begin{align}
    F(\dyad{\psi},\sigma)=\pi^m \int \dd\alpha W_{\dyad{\psi}}(\alpha) W_{\sigma}(\alpha),
\end{align}
while we only have blackbox access to $\sigma$.
We assume that we have full access or full knowledge of the Wigner function of $W_{\dyad{\psi}}(\alpha)$. This is a natural assumption in an experimental setting where the experimentalist wants to generate a specific quantum state.
Furthermore, we can sample from the Wigner function $W_{\sigma}(\alpha)$. 
In an experiment this would mean that the experimentalist performs a direct measurement of the Wigner function at point $\alpha$.
We can then estimate the fidelity in the following way.
We sample classically a point $\alpha$ according the probability distribution
\begin{align}
    p(\alpha)=\frac{\abs{\Tilde{W}_{\dyad{\psi}}(\alpha)}}{\norm{\Tilde{W}_{\dyad{\psi}}(\alpha)}_1}.
\end{align}
$\norm{\Tilde{W}_{\dyad{\psi}}(\alpha)}_1$ is the minimizer of the $\mu-$smoothed $l_1$-norm defined as 
\begin{align}
    \norm{W_{\dyad{\psi}}}_1^{(\mu)}= \inf_{\Tilde{W}_{\psi}, \norm{W_{\dyad{\psi}}-\Tilde{W}_{\psi}}_2\leq \mu} \norm{\Tilde{W}_{\psi}}_1.
\end{align}

For a chosen $\alpha$ we then obtain measure the Wigner function of state $\sigma$ at point $\alpha$ and obtain $W_{\sigma}(\alpha)$.
We obtain a sample $X=W_{\sigma}(\alpha) \norm{\Tilde{W}_{\dyad{\psi}}}_1 \text{sign}(\Tilde{W}_{\dyad{\psi}}(\alpha)) $.
Then by definition the expectation value of $X$ is the fidelity we want to estimate
\begin{align}
    \mathds{E}(X)&= \pi^m \int \dd\alpha \Tilde{W}_{\dyad{\psi}}(\alpha) W_{\sigma}(\alpha).
\end{align}
The number of samples needed to estimate the fidelity with error $\epsilon$ and failure probability $\delta$ is
\begin{align}
    N_{\epsilon,\delta}(\dyad{\psi})&\leq 2^{2m+1} \inf_{\mu \in [0,\epsilon)} \qty(\frac{\norm{W_{\dyad{\psi}}}_1^{(\mu)}}{\epsilon-\mu})^2\log(\frac{1}{\delta})\\ 
    &\leq 2^{2m+1}  \qty(\frac{\norm{W_{\dyad{\psi}}}_1}{\epsilon})^2\log(\frac{1}{\delta}).
\end{align}
We can use the standard Wigner function $W_{\dyad{\psi}}$ instead of the minimizer $\tilde{W}_{\dyad{\psi}}$ but are paying with a looser upper-bound and in consequence more samples to get guaranteed error $\epsilon$.
At the same time if we have a sample budget and sample $N$ times, we can upper bound the error meaning we can guarantee in which range the true fidelity lies.
The error for a one mode state can be upper bounded as
\begin{align}
    \epsilon\leq \norm{W_{\dyad{\psi}}}_1 \qty(2^3  N_{\epsilon,\delta}(\dyad{\psi})^{-1} \log(\frac{1}{\delta}))^{\frac{1}{2}}.
\end{align}

\subsection{Numerical sampling}
In this section we introduce how we implemented the protocol of the previous section.
We first introduce bins of the argument $\alpha$ of width $\gamma$ and consider the Wigner function of $W_\rho$ in these bins.
This is necessary to be able to transform the continuous probability distribution to a discrete one and sample numerically from it.

Then we sample $\alpha$ given probability 
\begin{align}
    p(\alpha)=\frac{\abs{W_{\rho}(\alpha)}}{\norm{W_{\rho}}_1}.
\end{align}
We sample from this distribution by performing inverse transform sampling.
Specifically, we first compute the marginal for $\mathcal{R}(\alpha)$ and obtain samples $\{\mathcal{R}(\alpha)_i\}$ using inverse transform sampling. Then we compute the conditional probabilities $p(\mathcal{I}(\alpha)|\mathcal{R}(\alpha)_i)$ and perform inverse transform sampling again to obtain $\mathcal{I}(\alpha)_i$. This gives us the samples $\mathcal{R}(\alpha)_i,\mathcal{I}(\alpha)_i$, which is just $\alpha_i$.
With this $\alpha_i$, we then use the experimental data and obtain the Wigner function $W_\sigma(\alpha_i)$. By repeating this $N$ times and choosing a satisfactory failure probability $\delta$, we then obtain the fidelity between the state one want and the state one has including a worst case error. 
The fidelity we are using the certify a certain stellar rank using stellar fidelity profiles uses these estimated fidelities minus $\epsilon/2$ as this is the lowest fidelity possible.

\section{Expression to compute the threshold values of the few-point witness}

In this section we prove Eq.(\ref{eq:pi_c}), %\cref{eq:threshold} 
and show how to compute threshold values for the few-points witnesses $\hat W(\bm a,\bm\alpha)$ introduced in \cref{sec:displaced_parity_wit}. 

\subsection{Brute-force computation}
\label{ap:proofII}

The threshold values can be computed as~\cite{Fiurasek:22}:
\begin{align}
    w_k(\bm a,\bm\alpha)=\max_{r^\star(\rho)\le n}\mathrm{Tr}[\hat W\rho]=\sup_{\hat G\in\mathcal G}[\mathrm{maxeig}(\hat\Pi_n \hat G \hat W \hat G^\dag \hat \Pi_n)].
\end{align}

We start with the simple case of a single-point witness, i.e.\ $\hat{W}(1,0)=\hat{\Pi}$. Then
\begin{align}
    \hat \Pi_k \hat G \hat \Pi \hat G^\dag \hat \Pi_k=\sum_{l=0}^{+\infty}(-1)^l\sum_{m,n=0}^k|m\rangle\langle m|\hat D(\beta)\hat S(\xi)|l\rangle\langle l|\hat S^\dag(\xi)\hat D^\dag(\beta)|n\rangle\langle n|.
\end{align}

Since the operators $\hat \Pi$ and $\hat \Pi_k$ are phase-invariant and since we are taking the supremum over all Gaussian unitary operations, we can take $\xi=r\in\mathbb R$ without loss of generality. Moreover, by looking at the Fock basis elements we have $\hat \Pi=\int_{\gamma\in\mathbb C}\hat D(\gamma)\frac{d^2\gamma}{2\pi}$. We thus obtain
\begin{align}
\hat S(r)\hat \Pi \hat S^\dag(r)&=\int_{\gamma\in\mathbb C}\hat S(r)\hat D(\gamma)S^\dag(r)\frac{d^2\gamma}{2\pi}\\
&=\int_{\gamma\in\mathbb C}\hat D(\cosh r\,\gamma+\sinh r\,\gamma^*)\frac{d^2\gamma}{2\pi}\\
&=\int_{\delta\in\mathbb C}\hat D(\delta)\frac{(\cosh^2r-\sinh^2r)d^2\delta}{2\pi}\\
&=\int_{\delta\in\mathbb C}\hat D(\delta)\frac{d^2\delta}{2\pi}\\
&=\hat\Pi.
\end{align}
The parity operator thus commutes with all squeezing operators so without loss of generality we can set $\hat G=\hat D(d)$, with $d\in\mathbb R$, due to the phase-invariance argument. We thus obtain
\begin{align}
\hat \Pi_k\hat G\hat \Pi\hat G^\dag\hat \Pi_k&=\hat \Pi_k\hat D(d)\hat \Pi \hat D^\dag(d)\hat \Pi_k\\
&=\hat \Pi_k\left(\int_{\gamma\in\mathbb C}\hat D(d)\hat D(\gamma)\hat D^\dag(d)\frac{d^2\gamma}{2\pi}\right)\hat \Pi_k\\
&=\hat \Pi_k\left(\int_{\gamma\in\mathbb C}e^{d(\gamma^*-\gamma)}\hat D(\gamma)\frac{d^2\gamma}{2\pi}\right)\hat \Pi_k\\
&=\sum_{0\le m,n\le k}\left(\int_{\gamma\in\mathbb C}e^{d(\gamma^*-\gamma)}\langle m|\hat D(\gamma)|n\rangle\frac{d^2\gamma}{2\pi}\right)|m\rangle\langle n|\\
&=:\sum_{0\le m,n\le k}\pi_{m,n}(d)|m\rangle\langle n|.
\end{align}
The Fock basis elements of a displacement operator are given by:
\begin{align}\label{eq:disp Fock basis}
   \langle m|D(\gamma)|n\rangle=e^{-\frac12|\gamma|^2}\sum_{p=0}^{\min m,n}\frac{\sqrt{m!n!}(-1)^{n-p}}{p!(m-p)!(n-p)!}\gamma^{m-p}\gamma^{*n-p}. 
\end{align}
This gives
\begin{align}
\pi_{m,n}(d)&=\int_{\gamma\in\mathbb C}e^{d(\gamma^*-\gamma)}\langle m|D(\gamma)|n\rangle\frac{d^2\gamma}{2\pi}\\
&=\sum_{p=0}^{\min m,n}\frac{\sqrt{m!n!}(-1)^{n-p}}{p!(m-p)!(n-p)!}\int_{\gamma\in\mathbb C}e^{-\frac12|\gamma|^2}e^{d(\gamma^*-\gamma)}\gamma^{m-p}\gamma^{*n-p}\frac{d^2\gamma}{2\pi}.
\end{align}

Let us rewrite this expression:
\begin{align}
&\int_{\gamma\in\mathbb C}e^{-\frac12|\gamma|^2}e^{d(\gamma^*-\gamma)}\gamma^{m-p}\gamma^{*n-p}\frac{d^2\gamma}{2\pi}=\int_{x,y\in\mathbb R}e^{-\frac12(x^2+y^2)}e^{-2iyd}(x+iy)^{m-p}(x-iy)^{n-p}\frac{dxdy}{2\pi}\\
&=\frac1{2\pi}i^m(-i)^n\sum_{s=0}^{m-p}\sum_{t=0}^{n-p}\binom{m-p}s\binom{n-p}t(-i)^si^t\int_{x\in\mathbb R}x^{s+t}e^{-\frac12x^2}dx\cdot\int_{y\in\mathbb R}y^{m+n-2p-s-t}e^{-2iyd}e^{-\frac12y^2}dy.
\end{align}

We have $\int_{x\in\mathbb R}x^qe^{-\frac12x^2}dx=\sqrt{2\pi}(q-1)!!$ ($0$ if $q$ is odd) and $\int_{y\in\mathbb R}e^{-2i\pi y\xi}y^qe^{-\frac12y^2}dy=\sqrt{2\pi}(-i)^q\mathrm{He}_q(2\pi\xi)e^{-2(\pi\xi)^2}$, where $\mathrm{He}$ are the Hermit polynomials. So we obtain

\begin{align}
\int_{\gamma\in\mathbb C}e^{-\frac12|\gamma|^2}e^{d(\gamma^*-\gamma)}\gamma^{m-p}\gamma^{*n-p}\frac{d^2\gamma}{2\pi}=(-1)^ne^{-2d^2}\!\!\sum_{\substack{0\le s\le m-p\\0\le t\le n-p\\s+t\text{ even}}}\binom{m-p}s\binom{n-p}t(-1)^{s+p}(s+t-1)!!\mathrm{He}_{m+n-2p-s-t}(2d).
\end{align}
This yields:
\begin{align}
&\pi_{m,n}(d)=e^{-2d^2}\sum_{p=0}^{\min m,n}\sum_{\substack{0\le s\le m-p\\0\le t\le n-p\\s+t\text{ even}}}\frac{\sqrt{m!n!}}{p!(m-p)!(n-p)!}\binom{m-p}s\binom{n-p}t(-1)^s(s+t-1)!!\mathrm{He}_{m+n-2p-s-t}(2d)\\
&=e^{-2d^2}\sum_{p=0}^{\min m,n}\sum_{\substack{0\le s\le m-p\\0\le t\le n-p\\s+t\text{ even}}}\frac{\sqrt{m!n!}(-1)^s(s+t)!}{2^{\frac{s+t}2}p!s!t!(m-p-s)!(n-p-t)!(\frac{s+t}2)!}\mathrm{He}_{m+n-2p-s-t}(2d)\\
&=e^{-2d^2}\sum_{p=0}^{\min m,n}\sum_{\substack{p\le u\le m\\p\le v\le n\\u+v\text{ even}}}\frac{\sqrt{m!n!}(-1)^{u-p}(u+v-2p)!}{2^{\frac{u+v}2-p}p!(u-p)!(v-p)!(m-u)!(n-v)!(\frac{u+v}2-p)!}\mathrm{He}_{m+n-u-v}(2d)\\
&=e^{-2d^2}\sum_{\substack{0\le u\le m\\0\le v\le n\\u+v\text{ even}}}\frac{\sqrt{m!n!}(-1)^u}{2^{\frac{u+v}2}(m-u)!(n-v)!}\mathrm{He}_{m+n-u-v}(2d)\sum_{p=0}^{\min u,v}\frac{(-1)^p2^p(u+v-2p)!}{p!(u-p)!(v-p)!(\frac{u+v}2-p)!}\\
&=e^{-2d^2}\sum_{\substack{0\le u\le m\\0\le v\le n\\u+v\text{ even}}}\frac{\sqrt{m!n!}(-1)^u}{2^{\frac{u+v}2}(m-u)!(n-v)!(\frac{u+v}2)!}\mathrm{He}_{m+n-u-v}(2d)\sum_{p=0}^{\min u,v}\binom{u+v-2p}{u-p}\binom{\frac{u+v}2}p(-1)^p2^p.
\end{align}

When $u\le v$ and $u+v=:2l$ is even we have
\begin{align}
\sum_{p=0}^{\min u,v}\binom{u+v-2p}{u-p}\binom{\frac{u+v}2}p(-1)^p2^p&=\sum_{p=0}^u\binom{2u+2l-2p}{u-p}\binom{u+l}p(-1)^p2^p\\
&=(-1)^u\sum_{q=0}^u\binom{2q+2l}q\binom{u+l}{q+l}(-1)^q2^{u-q}\\
&=\begin{cases}\binom{u+l}{\frac u2},\quad u\text{ even}\\0,\quad u\text{ odd}\end{cases}\\
&=\begin{cases}\binom{\frac{u+v}2}{\frac u2},\quad u\text{ even}\\0,\quad u\text{ odd}.\end{cases}
\end{align}
The last equation is also valid when $v\le u$ and $u+v$ is even. Hence, all the terms with $u$ or $v$ odd in the sum above vanish and we obtain
\begin{align}
\pi_{m,n}(d)&=e^{-2d^2}\sum_{\substack{0\le u\le m\\0\le v\le n\\u\text{ and }v\text{ even}}}\frac{\sqrt{m!n!}}{2^{\frac{u+v}2}(m-u)!(n-v)!(\frac u2)!(\frac v2)!}\mathrm{He}_{m+n-u-v}(2d)\\
&=e^{-2d^2}\sum_{a=0}^{\lfloor\frac m2\rfloor}\sum_{b=0}^{\lfloor\frac n2\rfloor}\frac{\sqrt{m!n!}}{2^a2^b(m-2a)!(n-2b)!a!b!}\mathrm{He}_{m+n-2a-2b}(2d).
\end{align}

Thus we can compute the threshold value for the witness $\hat{W}=\hat{\Pi}$ as
\begin{align}
    \max_{r^\star(\rho)\le n}\mathrm{Tr}[\hat W(1,0)\rho]=\sup_{\hat G\in\mathcal G}[\mathrm{maxeig}(\hat\Pi_n \hat G \hat \Pi \hat G^\dag \hat \Pi_n)]=\max_{d \in \mathbb{R}}\qty[\mathrm{maxeig}\qty(\sum_{0\le m,n\le k}\pi_{m,n}(d)|m\rangle\langle n|)].
\end{align}

\medskip

We now turn to the general case.
The threshold value for witnessing stellar rank $k$ is then given by
\begin{align}
    w_k(\bm a,\bm\alpha)=\max_{r^\star(\rho)\le n}\,\sum_{i=1}^Na_i\mathrm{Tr}[\hat D(\alpha_i)\hat \Pi \hat D^\dag(\alpha_i)\rho].
\end{align}

We have
\begin{align}
\hat \Pi_k\hat G\hat D(\alpha)\hat \Pi \hat D^\dag(\alpha)\hat G^\dag\hat \Pi_k&=\hat \Pi_k\hat D(\beta+\alpha\cosh r+\alpha^*\e^{\ii \phi}\sinh r)\hat\Pi \hat D^\dag(\beta+\alpha\cosh r+\alpha^*\e^{\ii \phi}\sinh r)\hat\Pi_k\\
&=\hat\Pi_k\left(\int_{\gamma\in\mathbb C}\hat D(\beta+\alpha\cosh r+\alpha^*\e^{\ii \phi}\sinh r)\hat D(\gamma)\hat D^\dag(\beta+\alpha\cosh r+\alpha^*\e^{\ii \phi}\sinh r)\frac{d^2\gamma}{2\pi}\right)\hat \Pi_k\\
&=\hat\Pi_k\left(\int_{\gamma\in\mathbb C}e^{(\beta+\alpha\cosh r+\alpha^*\e^{\ii \phi}\sinh r)\gamma^*-(\beta^*+\alpha^*\cosh r+\alpha \e^{-\ii \phi}\sinh r)\gamma)}\hat D(\gamma)\frac{d^2\gamma}{2\pi}\right)\hat \Pi_k\\
&=\sum_{0\le m,n\le k}\left(\int_{\gamma\in\mathbb C}e^{(\beta+\alpha\cosh r+\alpha^*\e^{\ii \phi}\sinh r)\gamma^*-(\beta^*+\alpha^*\cosh r+\alpha\e^{-\ii \phi}\sinh r)\gamma)}\langle m|\hat D(\gamma)|n\rangle\frac{d^2\gamma}{2\pi}\right)|m\rangle\langle n|\\
&=\sum_{0\le m,n\le k}\pi_{m,n}(\beta+\alpha\cosh r+\alpha^*\e^{\ii \phi}\sinh r)|m\rangle\langle n|,
\end{align}
where
\begin{align}
\pi_{m,n}(z)&=\int_{\gamma\in\mathbb C}e^{z\gamma^*-z^*\gamma}\langle m|\hat D(\gamma)|n\rangle\frac{d^2\gamma}{2\pi}\\
&=\sum_{p=0}^{\min m,n}\frac{\sqrt{m!n!}(-1)^{n-p}}{p!(m-p)!(n-p)!}\int_{\gamma\in\mathbb C}e^{-\frac12|\gamma|^2}e^{z\gamma^*-z^*\gamma}\gamma^{m-p}\gamma^{*n-p}\frac{d^2\gamma}{2\pi}.
\end{align}

Let us rewrite this expression: with $z:=a+ib$,
\begin{align}
&\int_{\gamma\in\mathbb C}e^{-\frac12|\gamma|^2}e^{z\gamma^*-z^*\gamma}\gamma^{m-p}\gamma^{*n-p}\frac{d^2\gamma}{2\pi}=\int_{x,y\in\mathbb R}e^{-\frac12(x^2+y^2)}e^{2ixb}e^{-2iya}(x+iy)^{m-p}(x-iy)^{n-p}\frac{dxdy}{2\pi}\\
&=\frac1{2\pi}i^m(-i)^n\sum_{s=0}^{m-p}\sum_{t=0}^{n-p}\binom{m-p}s\binom{n-p}t(-i)^si^t\int_{x\in\mathbb R}x^{s+t}e^{2ixb}e^{-\frac12x^2}dx\cdot\int_{y\in\mathbb R}y^{m+n-2p-s-t}e^{-2iya}e^{-\frac12y^2}dy.
\end{align}

We have $\int_{w\in\mathbb R}e^{-2i\pi w\xi}w^qe^{-\frac12w^2}dw=\sqrt{2\pi}(-i)^q\mathrm{He}_q(2\pi\xi)e^{-2(\pi\xi)^2}$ and $\mathrm{He}_q(-c)=(-1)^q\mathrm{He}_q(c)$, so we obtain

 \begin{equation}
    \int_{\gamma\in\mathbb C}e^{-\frac12|\gamma|^2}e^{z\gamma^*-z^*\gamma}\gamma^{m-p}\gamma^{*n-p}\frac{d^2\gamma}{2\pi}=(-1)^{n+p}e^{-2|z|^2}\sum_{\substack{0\le s\le m-p\\0\le t\le n-p}}\binom{m-p}s\binom{n-p}ti^{s-t}\mathrm{He}_{m+n-2p-s-t}(2a)\mathrm{He}_{s+t}(2b). 
 \end{equation}

This yields:

\begin{align}
\label{eq:pi}
 \pi_{m,n}(z)=e^{-2|z|^2}\sum_{p=0}^{\min m,n}\frac{\sqrt{m!n!}}{p!(m-p)!(n-p)!}\sum_{\substack{0\le s\le m-p\\0\le t\le n-p}}\binom{m-p}s\binom{n-p}ti^{s-t}\mathrm{He}_{m+n-2p-s-t}(2a)\mathrm{He}_{s+t}(2b).   
\end{align}
Thus we can compute the threshold value of the witness $\hat{W}(\bm a,\bm\alpha)=\sum_{i=1}^N a_i \hat{D}(\alpha_i)\hat \Pi \hat{D}(\alpha_i)^\dagger$  to witness stellar rank $k$ as
    \begin{align}
    \label{eq:threshold}
    w_k(\bm a,\bm\alpha)=\sup_{x,y\in\mathbb R,r\ge0,\phi \in [0,2\pi)]}\left[\mathrm{maxeig}\left(\sum_{j=1}^Na_j\sum_{0\le m,n\le k}\pi_{m,n}(x+iy+\alpha_j\cosh r+\alpha_j^*\e^{\ii \phi}\sinh r)|m\rangle\langle n|\right)\right].
\end{align}

\subsection{Compact expression}
\label{ap:compact}

 In this subsection we will derive a compact version of Eq.~\eqref{eq:pi}.
For $z=x+iy$ we define the two-variable exponential generating function as
\begin{equation}
\mathcal G(u,v;z)
:=
\sum_{m,n=0}^{\infty}
\frac{u^m v^n}{\sqrt{m!n!}}\,\pi_{m,n}(z).
\label{eq:def-generating-function}
\end{equation}
Equivalently, coefficient extraction gives
\begin{equation}
\frac{\pi_{m,n}(z)}{\sqrt{m!n!}}
=[u^m v^n]\,\mathcal G(u,v;z),
\label{eq:coefficient-definition}
\end{equation}
where $[u^m v^n]F$ denotes the coefficient of $u^m v^n$ in the
power-series expansion of $F$.  For later use, define
\begin{align}
K_{M,N}(x,y)
:={}&
\sum_{s=0}^{M}\sum_{t=0}^{N}
\binom{M}{s}\binom{N}{t}i^{s-t}
\mathrm{He}_{M+N-s-t}(2x)\mathrm{He}_{s+t}(2y).
\label{eq:def-KMN}
\end{align}
Then Eq.~\eqref{eq:pi} may be written as
\begin{equation}
\pi_{m,n}(z)
=e^{-2|z|^2}
\sum_{p=0}^{\min(m,n)}
\frac{\sqrt{m!n!}}{p!(m-p)!(n-p)!}
K_{m-p,n-p}(x,y).
\label{eq:correct-Hermite-K}
\end{equation}
Substituting Eq.~\eqref{eq:correct-Hermite-K} into
Eq.~\eqref{eq:def-generating-function} gives
\begin{align}
\mathcal G(u,v;z)
={}&
 e^{-2|z|^2}
 \sum_{m,n\geq0}u^m v^n
 \sum_{p=0}^{\min(m,n)}
 \frac{1}{p!(m-p)!(n-p)!}
 \notag\\
&\times
 \sum_{\substack{0\leq s\leq m-p\\0\leq t\leq n-p}}
 \binom{m-p}{s}\binom{n-p}{t}i^{s-t}
 \mathrm{He}_{m+n-2p-s-t}(2x)\mathrm{He}_{s+t}(2y).
\label{eq:G-substitution}
\end{align}

As a first step we want to remove the coupled summation limits.
Set $M=m-p$ and  $N=n-p$.
Then $m=M+p$, $n=N+p$, and the three indices $M,N,p$ range
independently over the nonnegative integers.  Hence
\begin{align}
\mathcal G(u,v;z)
={}&
 e^{-2|z|^2}
 \sum_{p=0}^{\infty}\frac{(uv)^p}{p!}
 \sum_{M,N\geq0}\frac{u^M v^N}{M!N!}
 \notag\\
&\times
 \sum_{s=0}^{M}\sum_{t=0}^{N}
 \binom{M}{s}\binom{N}{t}i^{s-t}
 \mathrm{He}_{M+N-s-t}(2x)\mathrm{He}_{s+t}(2y)
 \notag\\
={}&
 e^{-2|z|^2+uv}\,\mathcal H(u,v;x,y),
\label{eq:G-factor-p}
\end{align}
The factor $e^{uv}$ arises solely from
$\sum_{p\geq0}(uv)^p/p!$.  The remaining quantity is
\begin{align}
\mathcal H(u,v;x,y)
:={}&
 \sum_{M,N\geq0}\frac{u^M v^N}{M!N!}
 \sum_{s=0}^{M}\sum_{t=0}^{N}
 \binom{M}{s}\binom{N}{t}i^{s-t}
 \notag\\
&\times
 \mathrm{He}_{M+N-s-t}(2x)\mathrm{He}_{s+t}(2y).
\label{eq:def-H}
\end{align}

Using
\begin{equation}
 \frac{1}{M!}\binom{M}{s}=\frac{1}{s!(M-s)!},\quad\frac{1}{N!}\binom{N}{t}=\frac{1}{t!(N-t)!},
\end{equation}
and introducing $a=M-s$ and $b=N-t $, we can rewrite Eq.~\eqref{eq:def-H} as
\begin{align}
\mathcal H(u,v;x,y)
={}&
 \sum_{a,b,s,t\geq0}
 \frac{u^{a+s}v^{b+t}i^{s-t}}{a!b!s!t!}
 \mathrm{He}_{a+b}(2x)\mathrm{He}_{s+t}(2y)
 \notag\\
={}&
 \left[
 \sum_{a,b\geq0}\frac{u^a v^b}{a!b!}\mathrm{He}_{a+b}(2x)
 \right]
 \left[
 \sum_{s,t\geq0}\frac{(iu)^s(-iv)^t}{s!t!}\mathrm{He}_{s+t}(2y)
 \right].
\label{eq:H-factorization}
\end{align}
The probabilists' Hermite polynomials satisfy
\begin{equation}
 \sum_{k=0}^{\infty}\mathrm{He}_k(X)\frac{w^k}{k!}
 =e^{Xw-w^2/2}.
\label{eq:Hermite-generating}
\end{equation}
It holds that
\begin{align}
\sum_{a,b\geq0}\frac{A^aB^b}{a!b!}\mathrm{He}_{a+b}(X)
&=
 \sum_{k=0}^{\infty}\mathrm{He}_k(X)
 \sum_{a+b=k}\frac{A^aB^b}{a!b!}
 \notag\\
&=
 \sum_{k=0}^{\infty}\mathrm{He}_k(X)\frac{(A+B)^k}{k!}
 \notag\\
&=
 \exp\!\left[X(A+B)-\frac{(A+B)^2}{2}\right].
\label{eq:two-variable-Hermite}
\end{align}
Applying Eq.~\eqref{eq:two-variable-Hermite} to the first factor in
Eq.~\eqref{eq:H-factorization}, with $A=u$, $B=v$, and $X=2x$, gives
\begin{equation}
 \sum_{a,b\geq0}\frac{u^a v^b}{a!b!}\mathrm{He}_{a+b}(2x)
 =
 \exp\!\left[2x(u+v)-\frac{(u+v)^2}{2}\right].
\label{eq:x-Hermite-sum}
\end{equation}
For the second factor, take $A=iu$, $B=-iv$, and $X=2y$.  We obtain
\begin{equation}
 \sum_{s,t\geq0}\frac{(iu)^s(-iv)^t}{s!t!}\mathrm{He}_{s+t}(2y)
 =
 \exp\!\left[2iy(u-v)+\frac{(u-v)^2}{2}\right].
\label{eq:y-Hermite-sum}
\end{equation}
Combining Eqs.~\eqref{eq:x-Hermite-sum} and
\eqref{eq:y-Hermite-sum}, and using
\begin{equation}
 2x(u+v)+2iy(u-v)=2uz+2vz^*
\end{equation}
and
\begin{equation}
 -\frac{(u+v)^2}{2}+\frac{(u-v)^2}{2}=-2uv,
\end{equation}
we find
\begin{equation}
 \mathcal H(u,v;x,y)
 =
 e^{2uz+2vz^*-2uv}.
\label{eq:H-evaluated}
\end{equation}
Substitution into Eq.~\eqref{eq:G-factor-p} yields
\begin{equation}
\mathcal G(u,v;z)
=
\sum_{m,n=0}^{\infty}
\frac{u^m v^n}{\sqrt{m!n!}}\pi_{m,n}(z)
=
\exp\!\left[-2|z|^2-uv+2uz+2vz^*\right].
\label{eq:generating-function}
\end{equation}
Using this compact equation for the generating function, we only need to extract the correct coefficients to obtain$\pi_{m,n}(z)$. 
To extract the coefficient of $u^m v^n$, factorize the last exponential as
\begin{equation}
 e^{-uv+2uz+2vz^*}=e^{-uv}e^{2uz}e^{2vz^*}
\end{equation}
and Taylor .expand the three factors.
Multiplying the three series explicitly gives
\begin{equation}
 e^{-uv+2uz+2vz^*}
 =\sum_{p,a,b\geq0}
 \frac{(-1)^p(2z)^a(2z^*)^b}{p!a!b!}
 u^{p+a}v^{p+b}.
\label{eq:triple-series}
\end{equation}
The coefficient of $u^m v^n$ is therefore obtained precisely when
\begin{equation}
 p+a=m,\\
 p+b=n,
\end{equation}
so that $a=m-p$, $b=n-p$, and $0\leq p\leq\min(m,n)$.  Thus
\begin{align}
[u^m v^n]\,e^{-uv+2uz+2vz^*}
={}&\sum_{p=0}^{\min(m,n)}
\frac{(-1)^p(2z)^{m-p}(2z^*)^{n-p}}
{p!(m-p)!(n-p)!}.
\label{eq:coefficient-extracted}
\end{align}
Finally, using Eq.~\eqref{eq:coefficient-definition}, we obtain
\begin{equation}
\pi_{m,n}(z)
=
 e^{-2|z|^2}
 \sum_{p=0}^{\min(m,n)}
 \frac{\sqrt{m!n!}\,(-1)^p}
 {p!(m-p)!(n-p)!}
 (2z)^{m-p}(2z^*)^{n-p}.
\label{eq:compact-pi}
\end{equation}

%-------------------------------------------------------------------
\section{Derivation of odd- and even-parity cat state Wigner function under loss }\label{app:odd even Wigner}
In this appendix, we derive the Wigner functions of the odd- and even-parity cat states after the action of a loss channel. The corresponding density operators are
\begin{equation}
    \rho_{\alpha_0^{\pm}}=\frac{1}{\mathcal{N}^{(\pm)}}\left(\ketbra{\alpha_0}{\alpha_0}\pm\left[\ketbra{\alpha_0}{-\alpha_0}+\ketbra{-\alpha_0}{\alpha_0}\right]+\ketbra{-\alpha_0}{-\alpha_0}\right),
\end{equation}
where the normalization constant is given as 
\begin{align}
    \mathcal{N}^{(\pm)}=2\left(1\pm e^{-2|\alpha_0|^2}\right)
\end{align}
We model a loss channel $\Lambda_\mu(\cdot)$ as mixing a state with vacuum using a beam splitter from 
\begin{equation}
    \hat{BS}(\mu)=e^{\arccos(\sqrt{\mu})(\hat{a}_1^\dagger\hat{a}_2-\hat{a}_1\hat{a}_2^\dagger)}.
\end{equation}
The action of this beam splitter on two displacement operators is given as 
\begin{equation}
    \hat{BS}(\mu)\hat{D}(\alpha)\otimes\hat{D}(\beta)\hat{BS}^\dagger(\mu)=\hat{D}(\sqrt{\mu}\alpha+\sqrt{1-\mu}\beta)\otimes\hat{D}(-\sqrt{1-\mu}\alpha+\sqrt{\mu}\beta),
\end{equation}
which then follows that
\begin{equation}
    \Lambda_{\mu}(\ketbra{\alpha}{\beta})=\Tr_2\left[\hat{BS}(\mu)(\ketbra{\alpha}{\beta}\otimes\ketbra{0}{0})\hat{BS}^\dagger(\mu)\right]=e^{-(1-\mu)(|\alpha|^2+|\beta|^2-2\beta^*\alpha)/2}\ketbra{\sqrt{\mu}\alpha}{\sqrt{\mu}\beta}
\end{equation}
where we have used $\braket{\beta}{\alpha}=e^{-|\beta|^2/2-|\alpha|^2/2+\beta^*\alpha}$.

Applying this result to each term in the odd-parity cat state and using the linearity of $\Lambda_\mu$, we obtain the action of the loss channel on a odd-parity cat state as

\begin{align*}
    \Lambda_{\mu}(\rho_{\alpha_0^-}) & = \Tr_2\!\left[\hat{BS}(\mu)\left(\rho_{\alpha_0^-}\otimes\ketbra{0}{0}\right)\hat{BS}^\dagger(\mu)\right]\\
    & = \frac{1}{\mathcal{N}}\left(\Tr_2\!\left[\hat{BS}(\mu)\left(\ketbra{\alpha_0}{\alpha_0}\otimes\ketbra{0}{0}\right)\hat{BS}^\dagger(\mu)\right]-\Tr_2\!\left[\hat{BS}(\mu)\left(\ketbra{\alpha_0}{-\alpha_0}\otimes\ketbra{0}{0}\right)\hat{BS}^\dagger(\mu)\right]\right.\\
    & \quad\left.-\Tr_2\!\left[\hat{BS}(\mu)\left(\ketbra{-\alpha_0}{\alpha_0}\otimes\ketbra{0}{0}\right)\hat{BS}^\dagger(\mu)\right]+\Tr_2\!\left[\hat{BS}(\mu)\left(\ketbra{-\alpha_0}{-\alpha_0}\otimes\ketbra{0}{0}\right)\hat{BS}^\dagger(\mu)\right]\right)\\
    & = \frac{1}{\mathcal{N}}\left(\ketbra{\sqrt{\mu}\alpha_0}{\sqrt{\mu}\alpha_0}-e^{-2(1-\mu)|\alpha_0|^2}\left[\ketbra{\sqrt{\mu}\alpha_0}{-\sqrt{\mu}\alpha_0}+\ketbra{-\sqrt{\mu}\alpha_0}{\sqrt{\mu}\alpha_0}\right]+\ketbra{-\sqrt{\mu}\alpha_0}{-\sqrt{\mu}\alpha_0}\right)
\end{align*}
If we are assuming that $\alpha_0\in\mathbb{R}$, then we can derive the Wigner function of the odd-parity cat state under loss using the above relations 
\begin{align}
    W_{\Lambda_{\mu}(\rho_{\alpha_0^-})}(\alpha)& = \frac{2}{\pi}\Tr\left[\hat{D}(\alpha)\hat \Pi\hat{D}^\dagger(\alpha) \Lambda_{\mu}(\rho_{\alpha_0^-})\right] \\
    & =\frac{2}{\pi \mathcal{N}^{(-)}}\left(\bra{\sqrt{\mu}\alpha_0}\hat{D}(2\alpha)\hat{\Pi}\ket{\sqrt{\mu}\alpha_0}+\bra{-\sqrt{\mu}\alpha_0}\hat{D}(2\alpha)\hat{\Pi}\ket{-\sqrt{\mu}\alpha_0}\right.\\
    & \quad\left.-e^{-2(1-\mu)|\alpha_0|^2}\left[\bra{-\sqrt{\mu}\alpha_0}\hat{D}(2\alpha)\hat{\Pi}\ket{\sqrt{\mu}\alpha_0}+\bra{\sqrt{\mu}\alpha_0}\hat{D}(2\alpha)\hat{\Pi}\ket{-\sqrt{\mu}\alpha_0}\right]\right) \\
    & =\frac{2}{\pi \mathcal{N}^{(-)}}\left(\bra{\sqrt{\mu}\alpha_0}\hat{D}(2\alpha)\ket{-\sqrt{\mu}\alpha_0}+\bra{-\sqrt{\mu}\alpha_0}\hat{D}(2\alpha)\ket{\sqrt{\mu}\alpha_0}\right.\\
    & \quad\left.-e^{-2(1-\mu)|\alpha_0|^2}\left[\bra{-\sqrt{\mu}\alpha_0}\hat{D}(2\alpha)\ket{-\sqrt{\mu}\alpha_0}+\bra{\sqrt{\mu}\alpha_0}\hat{D}(2\alpha)\ket{\sqrt{\mu}\alpha_0}\right]\right) \\
    & =\frac{2}{\pi \mathcal{N}^{(-)}}\left(e^{-2|\alpha-\sqrt{\mu}\alpha_0|^2}+e^{-2|\alpha+\sqrt{\mu}\alpha_0|^2}-e^{-2(1-\mu)|\alpha_0|^2-2|\alpha|^2}\left[e^{4i\sqrt{\mu}\alpha_0\alpha_I}+e^{-4i\sqrt{\mu}\alpha_0\alpha_I}\right]\right) \\
    & =\frac{2}{\pi \mathcal{N}^{(-)}}\left(e^{-2[(\alpha_R-\sqrt{\mu}\alpha_0)^2+\alpha_I^2]}+e^{-2[(\alpha_R+\sqrt{\mu}\alpha_0)^2+\alpha_I^2]}-2e^{-2(1-\mu)|\alpha_0|^2-2(\alpha_R^2+\alpha_I^2)}\cos(4\sqrt{\mu}\alpha_0\alpha_I)\right) \\
    & =\frac{2}{\pi\mathcal{N}^{(-)}}\left(e^{-2(\alpha_R^2+\mu\alpha_0^2+\alpha_I^2)}\left[e^{4\sqrt{\mu}\alpha_0\alpha_R}+e^{-4\sqrt{\mu}\alpha_0\alpha_R}\right]-2e^{-2(1-\mu)|\alpha_0|^2-2(\alpha_R^2+\alpha_I^2)}\cos(4\sqrt{\mu}\alpha_0\alpha_I)\right) \\
    & =\frac{4}{\pi \mathcal{N}^{(-)}}e^{-2(\alpha_R^2+\alpha_I^2)}\left(e^{-2\mu\alpha_0^2}\cosh(4\sqrt{\mu}\alpha_0\alpha_R)-e^{-2(1-\mu)\alpha_0^2}\cos(4\sqrt{\mu}\alpha_0\alpha_I)\right)\label{eq:Wigner odd cat}
\end{align}
where we have used $\hat{\Pi}\hat{D}^\dagger(\alpha)=\hat{D}(\alpha)\hat{\Pi}$, $\hat{\Pi}\ket{\alpha}=\ket{-\alpha}$ and 
\begin{equation}
    \bra{\beta}\hat{D}(\alpha)\ket{\gamma} = e^{-|\alpha|^2/2-|\beta|^2/2-|\gamma|^2/2+\alpha\beta^*-\alpha^*\gamma+\beta^*\gamma},
\end{equation}
which can be derived by using the displacement decomposition $\hat{D}(\alpha)=e^{-|\alpha|^2/2}e^{\alpha\hat{a}^\dagger}e^{-\alpha^*\hat{a}}$. The Wigner function of the even-parity cat state is derived analogously and is given by
\begin{equation}\label{eq:Wigner even cat}
    W_{\Lambda_{\mu}(\rho_{\alpha_0^+})}(\alpha)=\frac{4}{\pi \mathcal{N}^{(+)}}e^{-2(\alpha_R^2+\alpha_I^2)}\left(e^{-2\mu\alpha_0^2}\cosh(4\sqrt{\mu}\alpha_0\alpha_R)+e^{-2(1-\mu)\alpha_0^2}\cos(4\sqrt{\mu}\alpha_0\alpha_I)\right).
\end{equation}
The Wigner functions of ideal odd- and even-parity cat states are obtained by setting $\mu=1$.
%--------------------------------------------------------------------
\section{Robustness of few-point witnesses to photon loss and displacement error}
\label{app:robustness}

This section aims to analyze the robustness of the multi-point and two-point witness to photon loss and displacement errors. For multi-point witness, we run simulations on the binomial state, the odd-parity cat state (magnitude $\sqrt{2}$), and on the physical GKP state, subject to photon loss error and displacement error, and analyze the robustness of the multi-points witness against these errors. In the case of loss robustness, we also compare the performance of the multi-points witness against the performance of the witness based on stellar profiles. For two-points witness, we study the range of Wigner function, which from Eq.~(\ref{eq:two_point_range}) is equal to the maximum value one can achieve using two-points witness (and hence gives an upper bound on the highest stellar rank we can witness) of odd and even-parity cat states of different magnitudes and observe how it behaved with photon loss and displacement error.
%----------------------------------------------------------------
\subsection{Robustness of few point witnesses to photon loss}\label{appsec:loss_photon}
Photon loss error is one of the most common sources of error in experiments, therefore it becomes paramount for a good witness of non-Gaussianity (here, stellar rank) to be robust against such losses. Here we test the robustness of the few-points witness to photon loss in two ways: In subsection \ref{subsec:multi_photon_loss}, we numerically compare the performance of the multi-point witness to stellar profile witness for the binomial state, odd-parity cat state with magnitude $\sqrt{2}$ and GKP state under photon loss. In subsection \ref{subsec:two_photon_loss}, we study the robustness of the two-point witness to photon loss for witnessing odd and even parity cat states by studying the decay of the range of wigner function for these two states with photon loss as from Eq.~(\ref{eq:two_point_range}) we know that the maximum obtainable value of witness (which determines the stellar rank we can witness) for a particular state is equal (up to a constant factor) to the range of the wigner function of the state.
%-------------------------------------------------------------------
\subsubsection{Robustness of multi-point witness to photon loss}\label{subsec:multi_photon_loss}
In Tables \ref{tab:stellar_photon_binomial},  \ref{tab:stellar_photon_cat}, and \ref{tab:stellar_photon_GKP}, we subject the binomial state, the odd-parity cat state with magnitude $\sqrt{2}$ and the physical GKP state respectively to photon loss $\eta$, and analyze how the stellar profile witness and few-points witness behaves with increasing photon losses (note that $\eta = 1$ represent no photon loss).

\begin{table*}[htbp]
    \centering
    \begin{tabular}{|c| c | c|}\hline
         $\eta$ & Stellar profile witness & Few-points witness\\\hline
        0.99 & 4 & 2 \\ \hline
        0.98 & 4 & 2 \\ \hline
        0.96 & 4 & 2 \\ \hline
        0.94 & 4 & 2 \\ \hline
        0.92 & 4 & 1\\ \hline
        0.9 & 3 & 1 \\ \hline
        0.85 & 2 & 1 \\ \hline
        0.8 & 2 & 1 \\ \hline
        0.75 & 1 & 0 \\ \hline
        0.7 & 0 & 0 \\ \hline
    \end{tabular}
    \caption{Stellar rank witnessed by stellar profile witness and few-points witness for the binomial state subject to photon loss. To compute the few point witness, we are using points I, II, III (Figure \ref{fig:bin_state}). We observe that stellar profile witness is more robust than few-points witness, giving non-zero stellar rank up to higher photon losses.}
    \label{tab:stellar_photon_binomial}
\end{table*}

\begin{table*}[htbp]
    \centering
    \begin{tabular}{|c| c | c|}\hline
         $\eta$ & Stellar profile witness & Few-points witness\\\hline
        0.99 & 2 & 3 \\ \hline
        0.98 & 1 & 2 \\ \hline
        0.96 & 1 & 2 \\ \hline
        0.94 & 1 & 1 \\ \hline
        0.9 & 1 & 1 \\ \hline
        0.85 & 1 & 1 \\ \hline
        0.8 & 1 & 0 \\ \hline
        0.7 & 1 & 0 \\ \hline
        0.6 & 1 & 0 \\ \hline
        0.55 & 0 & 0 \\ \hline
    \end{tabular}
    \caption{Stellar rank witnessed by stellar profile witness and few-points witness for the odd-parity cat state with magnitude $\sqrt{2}$ subject to photon loss. To compute the few point witness, we are using points I, II, III, IV, V (Figure \ref{fig:cat_state}). We observe that stellar profile witness is more robust than few-points witness, giving non-zero stellar rank up to higher photon losses, although for $\eta \geq 0.96$, the few point witness works better than the stellar profile witness.}
    \label{tab:stellar_photon_cat}
\end{table*}

\begin{table*}[htbp]
    \centering
    \begin{tabular}{|c| c | c|}\hline
         $\eta$ & Stellar profile witness & Few-points witness\\\hline
        0.995 & 11 &4\\ \hline
        0.99 & 10 & 4 \\ \hline
        0.985 & 8 & 1 \\ \hline
        0.98 & 8 & 1 \\ \hline
        0.96 & 6 & 1\\ \hline
        0.94 & 6 & 1 \\ \hline
        0.92 & 4 & 1 \\ \hline
        0.9 & 0 & 1 \\ \hline
        0.85 & 0 & 1 \\ \hline
        0.8 & 0 & 0 \\ \hline
    \end{tabular}
    \caption{Stellar rank witnessed by stellar profile witness and few-points witness for the GKP state subject to photon loss. To compute the few point witness, we are using points II, III, V (Figure \ref{fig:gkp_state}). We observe that few-points witness is more robust to photon loss, giving non-zero stellar rank up to higher photon losses.}
    \label{tab:stellar_photon_GKP}
\end{table*}

From the obtained data, we observe that stellar profile witnesses are more robust to photon losses than few-points witnesses for binomial and cat states, being able to witness non-zero stellar rank up to higher photon losses for these states. For GKP states however, few-point witness is more robust to photon losses than stellar profile witness. However, the exact value of $\eta$ up to which we can witness non-zero stellar rank depends on the state being observed, with both the stellar profile and the few point witnesses becoming trivial (giving zero) the fastest for physical GKP state. The stellar profile witness is more robust to photon losses for cat state then for the binomial state (stellar profile witness gives non-zero stellar rank up to $\eta =0.55$ for the cat state and up to $\eta =0.7$ for the binomial state), whereas the few-points witness is more robust to photon losses for the binomial state than the cat state (it gives non-zero stellar rank up to $\eta = 0.8$ for cat state and up to $\eta = 0.75$ for the binomial state). Finally, for extremely small photon loss (for $\eta$ as low as $0.96$), the few point witness works better than the stellar profile witness for the cat state. 

The difference in robustness of stellar profile witness and few-points witness reveals the trade-off one needs to think of when deciding whether to use stellar profile witness or few-points witness to witness the stellar rank of a given state since while few-points witness is more efficient in terms of the number of samples required, our analysis shows that the stellar profile witness can be more robust to photon losses.
%-------------------------------------------------------------------
\subsubsection{Robustness of two-point witness to photon loss}\label{subsec:two_photon_loss}

We investigate the ability of the two-point witness with equal-magnitude weights to certify the stellar rank of ideal odd- and even-parity cat states in the presence of loss. The Wigner functions of the odd and even-parity cat states under loss with real $\alpha_0$ are given in Eqs.~\eqref{eq:Wigner odd cat} and \eqref{eq:Wigner even cat}, respectively.
\begin{figure}[h]
    \centering
    \includegraphics[width=1.0\linewidth]{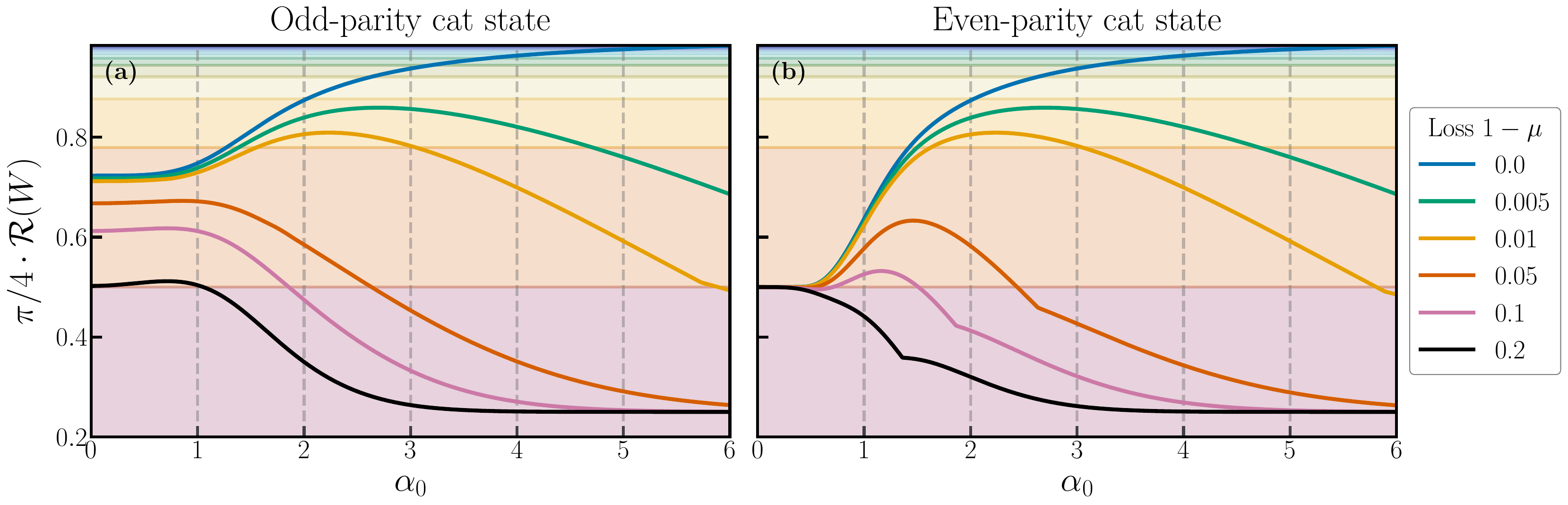}
    \caption{Panels (a) and (b) show the rescaled Wigner function ranges ($\pi/4 \cdot \mathcal{R}(W)$) of the ideal odd- and even-parity cat states, respectively, for five different loss levels. The shaded background regions indicate the intervals defined by the threshold values in Table~\ref{tab:two-point thresholds}, following the same color scheme as in Figure~\ref{fig:Fock odd cat ranges}.}
    \label{fig:loss cat}
\end{figure}

The ranges of the Wigner functions for odd- and even-parity cat states under different levels of loss are shown in Figures~\ref{fig:loss cat}(a) and \ref{fig:loss cat}(b), respectively. For both types of cat states, the Wigner-function range decreases rapidly even for small amounts of loss, substantially reducing the ability of the two-point witness to certify stellar rank. Moreover, the witness becomes increasingly sensitive to loss as $\alpha_0$ increases. This behavior is expected, since the average photon number grows with $\alpha_0$, making larger-amplitude cat states more susceptible to photon loss. Table~\ref{tab:largest stellar rank loss} summarizes the highest stellar rank that could be certified for lossy odd- and even-parity cat states.
\begin{table*}[htbp]
    \centering
\begin{tabular}{|c|c|c|}
    \cline{2-3}
    \multicolumn{1}{c|}{} &
    \multicolumn{2}{c|}{Largest certifiable stellar rank} \\
    \hline
    $\eta$ & Odd-cat state & Even-cat state \\ \hline
    0.9995 & 2 & 2 \\ \hline
    0.99   & 2 & 2 \\ \hline
    0.95   & 1 & 1 \\ \hline
    0.9    & 1 & 1 \\ \hline
    0.8    & 1 & 0 \\ \hline
\end{tabular}
    \caption{The maximum stellar rank witnessed by two-points witness for the odd- and even-cat states state subject to photon loss for $0\leq\alpha_0\leq6$.}
    \label{tab:largest stellar rank loss}
\end{table*}

%--------------------------------------------------------------------
\subsection{Displacement robustness of few-points witness}\label{appsec:loss_displacement}
In this section, we analyze the robustness of the multi-point and two-point witness to three different types of error: measurement error due to the imprecision in measurements, displacement error on the state and if we incorrectly pick the Wigner function points. For multi-point witness (Subsection \ref{subsec:multi_displacement}), we  analyze the robustness of both the theoretical and experimentally obtained binomial state, odd-parity cat state (magnitude $\sqrt{2}$) and physical GKP state if, instead of computing the witness value at the desired points on the Wigner function of the state, we compute the value of witness at points either linearly or randomly displaced from the desired witness. Similarly, for two-point witness (Subsection \ref{subsec:two_displacement}), we analyze the stellar rank witnessed for odd and even-parity cat states of different magnitudes when instead of computing the witness at the maxima and minima of the Wigner function, we compute it at a random displacement away.
%-----------------------------------------------------------------
\subsubsection{Robustness of multi-point witness to displacement error}\label{subsec:multi_displacement}
To compute the robustness of multi-point witness to displacement error, we first analyze how the stellar rank that can be witnessed for binomial state, odd-parity cat state (magnitude $\sqrt{2}$) and GKP state change when the points at which the Wigner function is probed are linearly displaced by $\epsilon$ in the positive $x$ direction from the desired point (Tables \ref{tab:stellar_displacement_binomial}, \ref{tab:stellar_displacement_cat}, \ref{tab:stellar_displacement_GKP}). This also gives an intuition on how the robustness of the multi-point witness to displacement error depends on the Wigner function of the state being probed.
\begin{table*}[htbp]
    \centering
    \begin{tabular}{|c| C{3.5 cm} | C{3.5cm}|}\hline
         $\epsilon$ & Few point witness value (theoretical state) & Stellar rank witnessed (theoretical state)  \\\hline
        0.0 & 0.6441 & 2 \\ \hline
        0.05 & 0.6314 &2 \\ \hline
        0.1 & 0.59315 & 2 \\ \hline
        0.15 & 0.53301 & 2  \\ \hline
        0.2 & 0.45655 & 1 \\ \hline
        0.25 & 0.3705 &1 \\ \hline
        0.3 & 0.2820 & 0 \\ \hline
    \end{tabular}
    \caption{Stellar rank witnessed for the ideal and experimentally obtained binomial state, when instead of computing the witness value at points I, II, III (Figure \ref{fig:bin_state}) , we are instead computing the witness by picking the points at $\epsilon$ distance from these points. We observe that upto values of $\epsilon$ up to 0.3 for both the ideal state and the experimental state, the few point witness gives a non-zero stellar rank, making the few point witness somewhat robust to displacement error for the binomial state. }
    \label{tab:stellar_displacement_binomial}
\end{table*}

\begin{table*}[htbp]
    \centering
    \begin{tabular}{|c| C{3.5 cm} | C{3.5cm}|C{3.5cm} | C{3.5 cm}|}\hline
         $\epsilon$ & Few point witness value (theoretical state) & Stellar rank witnessed (theoretical state) \\\hline
        0.0 & 0.6368 & 3  \\ \hline
        0.1 & 0.6236 &3 \\ \hline
        0.15 & 0.60735 & 2  \\ \hline
        0.2 & 0.58501 & 2   \\ \hline
        0.25 & 0.5572 & 2 \\ \hline
        0.3 & 0.52464 & 1  \\ \hline
        0.35 & 0.48805 &1 \\ \hline
        0.4 & 0.44825 & 1  \\ \hline
        0.45 & 0.40605 & 1 \\ \hline
        0.5 & 0.3622 & 0  \\ \hline
    \end{tabular}
    \caption{Stellar rank witnessed for the ideal and experimentally obtained cat state, when instead of computing the witness value at points I, II, III, IV, V (Figure \ref{fig:cat_state}), we are instead computing the witness by picking the points at $\epsilon$ distance from these points. We observe that upto relatively high values of $\epsilon$ up to 0.5 for the ideal state and 0.45 for the experimental state, the few point witness gives a non-zero stellar rank, making the few point witness robust to displacement error for the cat state.}
    \label{tab:stellar_displacement_cat}
\end{table*}

\begin{table*}[htbp]
    \centering
    \begin{tabular}{|c| C{3.5 cm} | C{3.5cm}|}\hline
         $\epsilon$ & Few point witness value (theoretical state) & Stellar rank witnessed (theoretical state)  \\\hline
        0.0 & 0.73532 & 5  \\ \hline
        0.025 & 0.72788 &4 \\ \hline
        0.05 & 0.706003 & 4  \\ \hline
        0.075 & 0.671022 & 3  \\ \hline
        0.1 & 0.62504 & 1 \\ \hline
        0.2 & 0.38605 & 1 \\ \hline
        0.25 & 0.271 & 0 \\ \hline
    \end{tabular}
    \caption{Stellar rank witnessed for the ideal and experimentally obtained GKP state, when instead of computing the witness value at points II, III, V (Figure \ref{fig:gkp_state}), we are instead computing the witness by picking the points at $\epsilon$ distance from these points. We observe that for $\epsilon$ up to 0.25 for the ideal and experimental GKP state, the few point witness gives a non-zero stellar rank, making the few point witness not robust to displacement error for the GKP state. }
    \label{tab:stellar_displacement_GKP}
\end{table*}

We observe that the robustness of the few-points witness to displacement error on the state depends on the nature of Wigner function of the state whose stellar rank we are trying to witness. If the Wigner function of the state being probed is smooth around its maxima and minima, like it is the case for the cat state (Figure \ref{fig:cat_state}), we can witness a non-zero stellar rank up to relatively large values of $\epsilon$ (up to $\epsilon = 0.5$ for the theoretical cat state and $\epsilon = 0.45$ for the experimental cat state), whereas if the Wigner function of the state being probed changes sharply around its maxima/minima, like it is the case for the physical GKP state (Figure \ref{fig:gkp_state}), we can witness a non-zero stellar rank up to extremely small values of $\epsilon$ (up to $\epsilon = 0.25$ for theoretical and experimental physical GKP state). Since the sharpness of the binomial state around its maxima/minima is somewhere in between that of the cat state and the physical GKP state (Figure \ref{fig:gkp_state}), the robustness of the fidelity witness to displacement error falls somewhere in between as well (up to $\epsilon = 0.3$ for both the theoretical and experimental binomial state).

To further investigate the robustness of the witness against imperfect displacement operations, we perturb the phase-space sampling points $\alpha_j$ by adding 1000 random displacement vectors drawn from a two-dimensional Gaussian distribution centered at the $\alpha_j$ with standard deviation $\sigma_\alpha$. This provides a numerical model for displacement uncertainty and is motivated by the experimental implementation, where the protocol was shown to be robust against variations in the displacement pulse amplitude \cite{Kudra}.

\begin{figure}[H]
    \centering
    \includegraphics[width=0.65\textwidth]{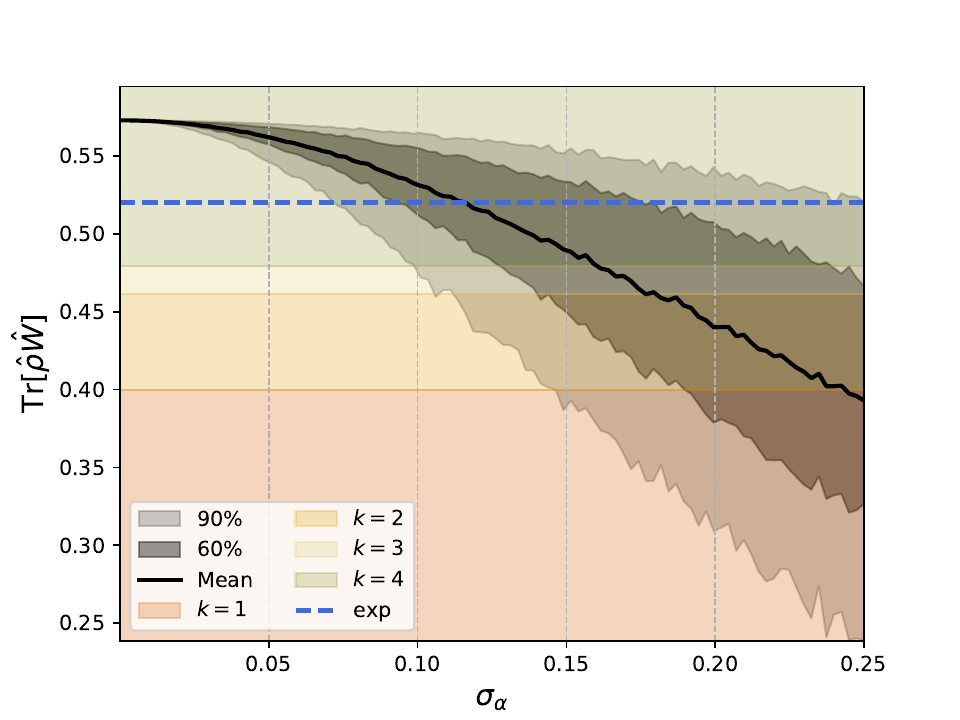}
    \caption{Displacement robustness analysis, for the ideal (black) and experimentally (blue) obtained binomial state with the witness constructed by the points I, II, III, IV, V (Figure \ref{fig:bin_state}). The thresholds were obtained from the non-displaced points as in table \ref{tab:thresh_values_bin} and the colored regions mark the corresponding stellar rank $k$. We compute the witness measurement $\text{Tr}[\rho\hat{W}]$, instead with 1000 samples of a Gaussian distribution with mean zero displacement and standard deviation $\sigma_\alpha$ for all five points.}
    \label{fig:displacement_robustness_bin}
\end{figure}

\begin{figure}[H]
    \centering
    \includegraphics[width=0.65\textwidth]{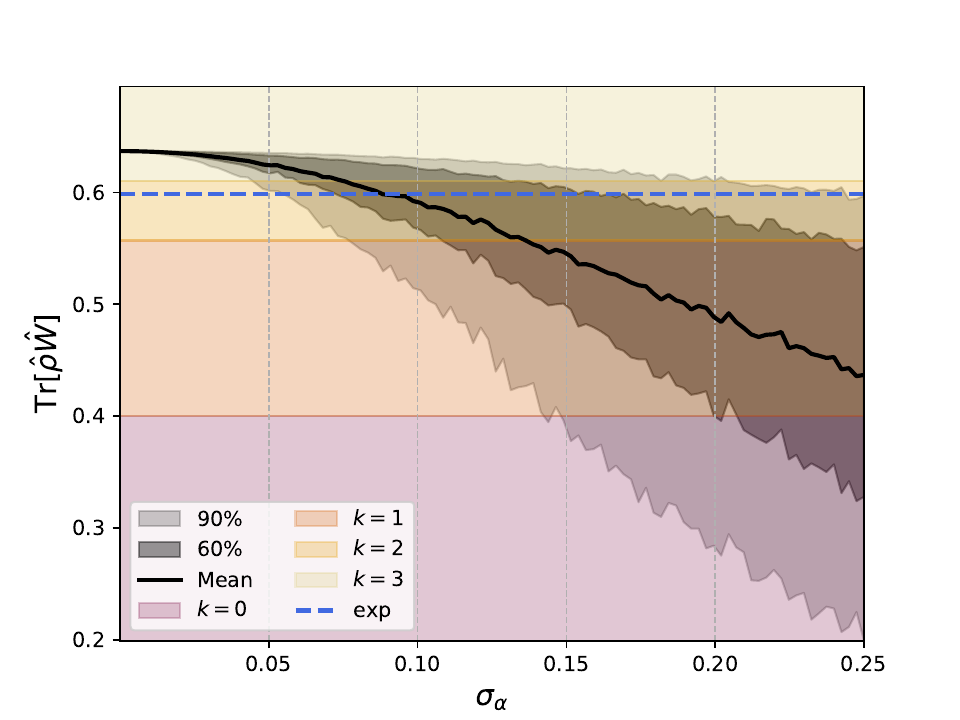}
    \caption{Displacement robustness analysis, for the ideal (black) and experimentally (blue) obtained cat state with the witness constructed by the points I, II, III, IV, V (Figure \ref{fig:cat_state}). The thresholds were obtained from the non-displaced points as in table \ref{tab:thresh_values_cat} and the colored regions mark the corresponding stellar rank $k$. We compute the witness measurement $\text{Tr}[\rho\hat{W}]$, instead with 1000 samples of a Gaussian distribution with mean zero displacement and standard deviation $\sigma_\alpha$ for all five points.}
    \label{fig:displacement_robustness_cat}
\end{figure}

\begin{figure}[H]
    \centering
    \includegraphics[width=0.65\textwidth]{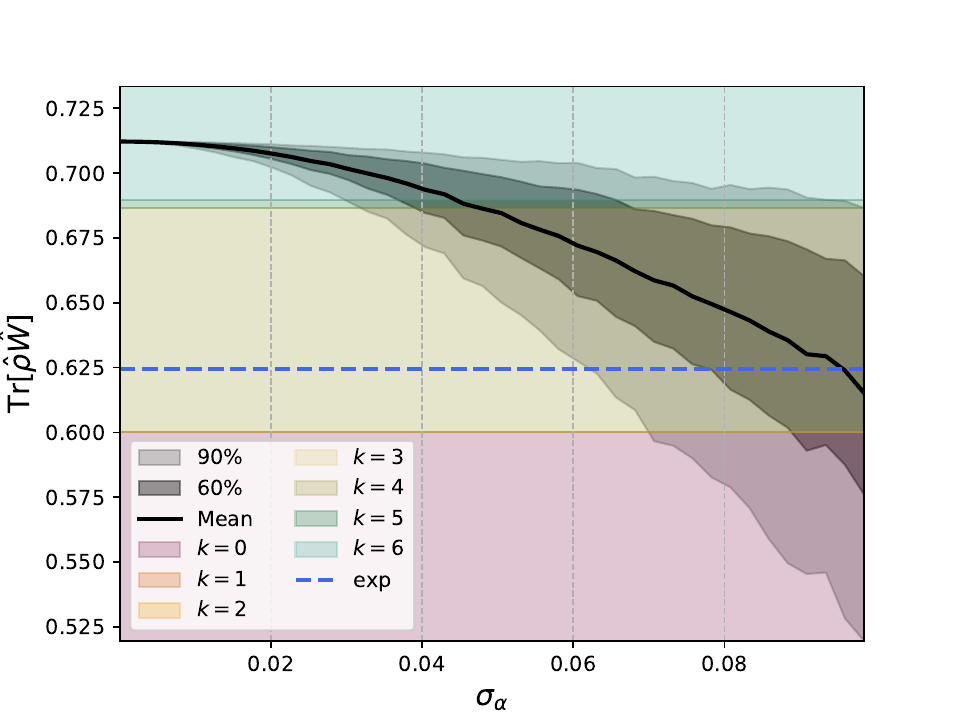}
    \caption{Displacement robustness analysis, for the ideal (black) and experimentally (blue) obtained GKP state with the witness constructed by the points I, II, III, IV, V (Figure \ref{fig:gkp_state}). The thresholds were obtained from the non-displaced points as in table \ref{tab:thresh_values_gkp} and the colored regions mark the corresponding stellar rank $k$. We compute the witness measurement $\text{Tr}[\rho\hat{W}]$, instead with 1000 samples of a Gaussian distribution with mean zero displacement and standard deviation $\sigma_\alpha$ for all five points.}
    \label{fig:displacement_robustness_gkp}
\end{figure}

The robustness analysis of the few points witnesses is performed using ideal theoretical Wigner functions,  while the witness obtained from the unperturbed experimental Wigner functions is shown as the blue reference curve in Figures~\ref{fig:displacement_robustness_bin}, \ref{fig:displacement_robustness_cat} and \ref{fig:displacement_robustness_gkp}.   % rather than the experimental data. Since the experimentally reconstructed Wigner functions already include any displacement imperfections introduced during the tomography, these contributions cannot be isolated and varied independently. Consequently, the robustness analysis is carried out on theoretical data,

The results confirm the robustness observed in the previous stability analysis while extending it to isotropic displacement perturbations in phase space. Compared with the one-dimensional displacement perturbation $\epsilon$ along the (x)-axis, the witness remains robust against displacement errors from arbitrary directions. The perturbation strength was varied up to $\sigma_\alpha = 0.25$ for the binomial and the cat state and to $\sigma_\alpha=0.1$ for the GKP state, because beyond these limits all investigated states are likely to yield a trivial certified stellar rank of zero. Interestingly, the GKP state appears to be more sensitive to the multidirectional displacement, for higher $\sigma_\alpha$ compared to the other to states, which might be due to the close thresholds of the witness. For the cat state we also observe a higher sensitivity than before in the linear displacement analysis, due to the close extrema in the interference region of the cat state along the imaginary axis. However, since in total five points contribute to the witness and the threshold bands are larger, the robustness of the cat witness exceeds the one of the GKP witness. 

%------------------------------------------------------------------

% \begin{figure}[h]
%     \centering
%     \includegraphics[width=1.0\linewidth]{Odd_even_range_disp_error.pdf}
%     \caption{Figures (a) and (b) show the ranges (rescaled with $\pi/4$) of the odd- and even-parity cat states, respectively, for five different losses. Figures (c) and (d) shows the displacement robustness analysis for odd- and even-parity cat states, respectively. The shaded regions indicate the fractions of witness expectation values obtained from evaluations at points randomly displaced from the ideal maximum and minimum locations for $10\,000$ independent samples. The displacements were drawn from a normal distribution with zero mean and standard deviation $\sigma = 0.01$.}

%     \label{fig:loss disp cat}
% \end{figure}

\subsubsection{Robustness of two-point witness to displacement error}\label{subsec:two_displacement}
\begin{figure}[h]
    \centering
    \includegraphics[width=1.0\linewidth]{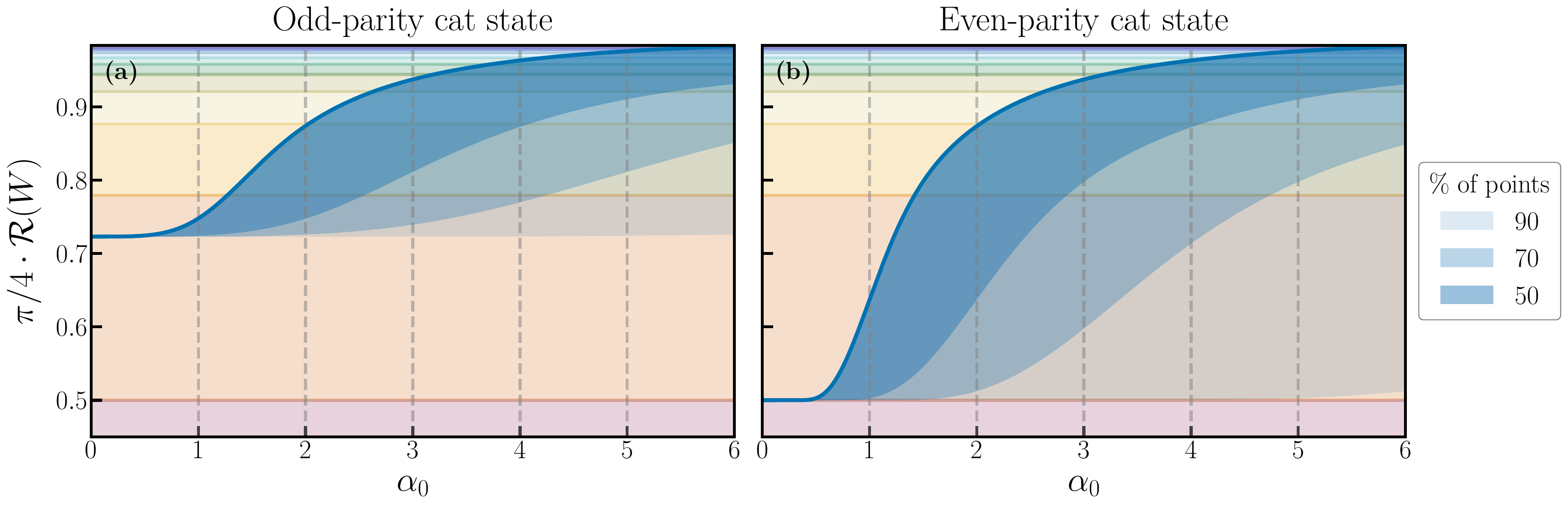}
    \caption{Panels (a) and (b) show the robustness of the rescaled Wigner function ranges ($\pi/4 \cdot \mathcal{R}(W)$) to displacement errors for ideal odd- and even-parity cat states, respectively. The shaded regions indicate the ranges containing 90\%, 70\%, and 50\% of the witness expectation values obtained from $10\,000$ independent samples, evaluated at randomly displaced maximum and minimum locations. The displacement errors are sampled from a zero-mean Gaussian distribution with standard deviation $\sigma=0.01$. The shaded background regions indicate the intervals defined by the threshold values in Table~\ref{tab:two-point thresholds}, following the same color scheme as in Figure~\ref{fig:Fock odd cat ranges}.}

    \label{fig:disp error cat}
\end{figure}

In this analysis for two-point witness, we apply independent random displacements to the phase-space points corresponding to the ideal maximum and minimum values of the witness and we then evaluate whether the corresponding stellar rank can still be certified. The results are shown in Figures~\ref{fig:disp error cat}(a) and \ref{fig:disp error cat}(b).

The cat states are again found to be  sensitive to this type of error. For both even- and odd-parity cat states, the maximum stellar rank that can be certified in at least $50\%$ of the trials is at most $4$, which is only possible for large amplitudes. This behavior is expected, since the extrema of the Wigner function are located within the interference fringes, where the phase-space structure is particularly sharp. Consequently, even small displacement errors can lead to significant changes in the Wigner function values at the sampled points, resulting in a noticeable degradation of the witness.

On the other hand, for smaller values of $\alpha_0$, the stellar rank witnessed is less sensitive to displacement errors. This is because the corresponding Wigner functions in general exhibit smoother features, causing the witness values to vary less rapidly under small displacements.
%----------------------------------------------------
\section{Witnessing stellar rank of $\ket{7}$ using few-points witness}\label{appsec:fock7_few_points}
To witness the stellar rank of $\ket{7}$, we restrict to the two-point witness by computing the witness at the minima and maxima of the Wigner function and investigate whether the certification of stellar rank 7 is theoretically achievable using an ideal Fock state $\ket{7}$. Note that none of the experimentally relevant states previously analyzed with the few point witnesses allowed us to certify stellar rank 7.

The radii of the witness points are fixed to the minima and maxima of the Wigner function, while the phases are optimized by considering different point configurations. Since the computational cost of threshold optimization increases rapidly with both the number of witness points and the threshold order $k$, identifying efficient point configurations is of practical importance.

\begin{figure*}[t]
\centering
\begin{minipage}{0.33\textwidth}
    \centering
    \includegraphics[width=\linewidth]{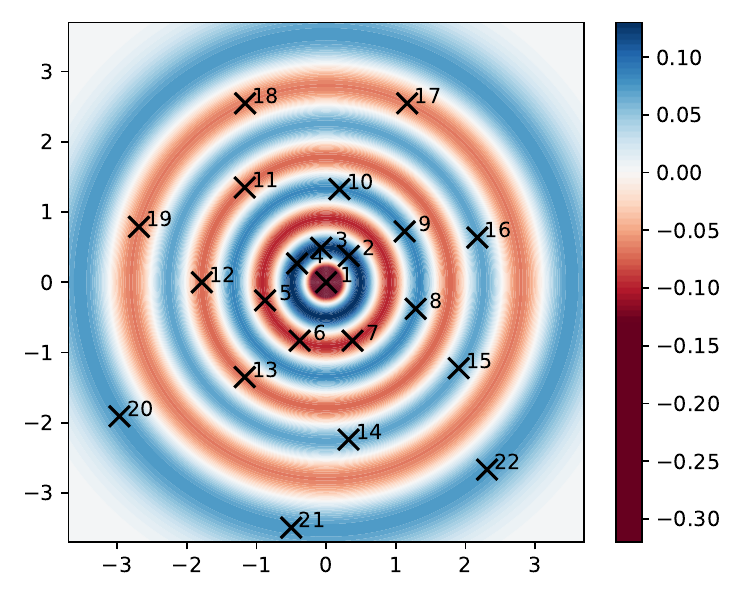}
    \caption{Wigner function of an ideal Fock state $\ket{7}$. The crosses indicate the location of points of the Wigner functions used to define the corresponding few-point witnesses.}
    \label{fig:fock7_state}
\end{minipage}
\hfill
\begin{minipage}{0.65\textwidth}
    \centering
\begin{tabular}{c| c c c c  | c }
        $k$ & 4 & 5 & 6 & 7 & Th.\\\hline
        
        Equal phase $\varphi\in\{0,\frac{2}{3}\pi, \frac{4}{3}\pi\}$ & 0.2617 & 0.2654 & 0.2996 & 0.3217 & 0.2951 \\
        Random phases & 0.2601 & 0.2764 & 0.2961 & 0.3572 & 0.2951 \\
        Spiral       & 0.2236 & 0.2399 & 0.2532 & 0.3069 & 0.2951 \\
   
    \end{tabular}
    \captionof{table}{Table of thresholds values $w_k$ for $\ket{7}$ and few-points witness expectation values. Each line corresponds to a few-point witness of 22 points (3 per radius except one at the origin) with a different phase configuration. Note: the "random-phase" configuration in the second line for the few-point witness considered here should not be confused with the phase-averaged witness discussed in a later subsection of the appendix, as well as in the main text. The spiral configuration is shown in Figure~\ref{fig:fock7_state}. An expectation value (Th.) higher than a threshold value $w_k$ certifies a stellar rank $>k$. The theoretical measurement values are the same because of the rotation symmetry of the Fock state and the construction choice of keeping the same radii values. 
    Only the spiral configuration certifies a correct stellar rank of $\ge 7$, the other two certify only a stellar rank of $\ge 6$.}
    \label{tab:thresh_values_fock7}
\end{minipage}
\end{figure*}

Among the tested configurations, a spiral arrangement of the witness points performs best. It certifies the correct stellar rank of the ideal Fock state using only 22 witness points (three points per radius) while providing the lowest thresholds for high $k$, thereby increasing robustness against experimental imperfections. The corresponding results are summarized in Table~\ref{tab:thresh_values_fock7}. Since we analyze an ideal Fock state and vary only the phase distribution of the point sets, the measurement values remain the same because of the rotation symmetry. Therefore, the witness with the lowest threshold values is the most desirable. 

It is noteworthy that the choice of the point phases has an influence on the witness performance, despite the rotational symmetry of the Fock state. This demonstrates that the geometric arrangement of the phase-space points constitutes a non-trivial optimization problem and that substantial improvements can be achieved even without changing the witness radii. %In section , we compare the performance of the 22-point witness to phase-averaged witness for witnessing stellar rank of $\ket{7}$.
% 
%-------------------------------------------------------------------------------------------
\section{Phase-averaged witness and its properties}\label{appsec:phase_averaged_witness}
\subsection{Phase averaging} \label{app:pa_witness_phase_averaging}
For a general mixed state, we can compute the phase averaging and derive some properties that might be helpful for the computation of a phase-averaged witness. We use the fact that the Fock states are eigenstates of the number operator $\hat{n}$, which allows us to write: 
\begin{align}
    \rho^{PA} &= \int_0^{2\pi}\frac{d\theta}{2\pi} \hat{R}(\theta) \rho \hat{R}^\dagger(\theta) \\
    &= \int_0^{2\pi}\frac{d\theta}{2\pi} \bigg(\sum^\infty_{n,m=0} \rho_{nm} e^{i\theta\hat{n}} \ket{n}\bra{m} e^{-i\theta\hat{n}} \bigg) \\
    &= \sum^\infty_{n,m=0} \rho_{nm} \int_0^{2\pi}\frac{d\theta}{2\pi} e^{i\theta n}  \ket{n}\bra{m} e^{-i\theta m}  \\
    &= \sum^\infty_{n,m=0} \rho_{nm} \ket{n}\bra{m} \underbrace{\int_0^{2\pi}\frac{d\theta}{2\pi} e^{i\theta (n-m)}}_{=\delta_{nm}} \\
    &= \sum^\infty_{n=0} \rho_{nn} \ket{n}\bra{n}.
\end{align}
The phase average of an arbitrary state consists only of the diagonal elements of that state, and the others are zero. A state is invariant under phase-space rotations if and only if its density operator is diagonal in the Fock basis. 

If one considers the optimization problem in Eq.~\eqref{eq:optimization_problem}, only over rotation-symmetric states, the resulting thresholds would correspond to the expression of a phase-averaged witness, 
\begin{align}
    w^{PA}_k(\boldsymbol{a},\boldsymbol{\alpha}) = \sup_{r^*(\rho)\le k} \text{Tr} \big( \rho^{PA} \hat{W}(\boldsymbol{a},\boldsymbol{\alpha}) \big).
\end{align}
The witness measurement could be redefined through a phase-averaging of the input state and a consecutive measurement of the Wigner function at the points/radii, 
\begin{align}
    \hat{W}_{PA}(\boldsymbol{a}, \boldsymbol{r}) = \mathcal{E}^*_{PA}\big( \hat{W}(\boldsymbol{a}, \boldsymbol{r}) \big).
\end{align}

%\begin{itemize}
    %\item image of state
    %\item mention phase randomized coherent states for quantum teleportation
    %\item stellar rank categorization lowered by phase-averaging
    %\item possibly more stable/robust certification 
    %\item why we did not pursue this approach
%\end{itemize}

\subsection{Properties of the witness} \label{app:pa_witness_properties}

Since Fock states are rotationally symmetric, in this section, we aim to construct a rotationally symmetric witness that uses the phase average for its construction instead of single points of the Wigner function. 
Here, a circle around the origin is evaluated for the Wigner function,
\begin{align}
    \hat{W}_{PA}(\boldsymbol{a}, \boldsymbol{r}) 
    &= \sum_{j=1}^N a_j \int_0^{2 \pi} \frac{d \theta}{2 \pi} \hat{D}(\alpha_j) \hat{\Pi} \hat{D}^\dagger(\alpha_j) \bigg|_{|\alpha_j|=r_j} \\
    &= \sum_{j=1}^N a_j \int_0^{2 \pi} \frac{d \theta}{2 \pi} \hat{R}(\theta) \hat{D}(r_j) \hat{R}^\dagger(\theta) \hat{\Pi} \hat{R}(\theta) \hat{D}^\dagger(r_j) \hat{R}^\dagger(\theta) \\
    &= \sum_{j=1}^N a_j \int_0^{2 \pi} \frac{d \theta}{2 \pi} \hat{R}(\theta) \hat{D}(r_j) \hat{\Pi} \hat{D}^\dagger(r_j) \hat{R}^\dagger(\theta).
\end{align}
In the second line, we use:
\begin{align*}
    \hat{D}(\alpha_j) 
    &= \exp(\alpha_j \hat{a}^\dagger - \alpha_j^* \hat{a}) 
    \overset{\alpha_j=r_je^{i\theta}}{=} \exp(r_je^{i\theta} \hat{a}^\dagger - r_je^{-i\theta} \hat{a}) \\
    &= \hat{R}(\theta) \exp(r_j \hat{a}^\dagger - r_j \hat{a}) \hat{R}^\dagger(\theta)
\end{align*}
and in the third line, the commutation relation of the parity operator with the rotation operator.
Using this construction of the witness, we can show it is rotation invariant and that the representation in the symmetric Fock basis is diagonal,
\begin{align}
    W_{PA}^{(nm)}(a_1, r_1) 
    &= \langle n | \hat{W}_{PA}(a_1, r_1) | m \rangle
    = a_1 \langle n | \int_0^{2 \pi} \frac{d \theta}{2 \pi} \hat{R}(\theta) \hat{D}(r_1) \hat{\Pi} \hat{D}^\dagger(r_1) \hat{R}^\dagger(\theta) | m \rangle \\
    &= a_1 \int_0^{2 \pi} \frac{d \theta}{2 \pi} \langle n | e^{i\theta n} \hat{D}(r_1) \hat{\Pi} \hat{D}^\dagger(r_1) e^{-i\theta m} | m \rangle \\
    &= a_1 \langle n | \hat{D}(r_1) \hat{\Pi} \hat{D}^\dagger(r_1)  | m \rangle \int_0^{2 \pi} \frac{d \theta}{2 \pi} e^{i\theta (n-m)} \\
    &= a_1 \langle n | \hat{D}(r_1) \hat{\Pi} \hat{D}^\dagger(r_1)  | m \rangle \delta_{nm}.
\end{align}
Here we only consider a one-point/one-radius witness, since the phase integral is a linear operator, and the result of the one-point witness can easily be generalized to the sum of multiple witness elements as in the few-point case. 
We will use the same trick to compute the general expression for the diagonal elements with respect to the chosen radius and formulate the one-point threshold optimization problem first and then write down the general expression afterward. 
For the diagonal one-point witness elements, we get the following:
\begin{align}
    W_{PA}^{(nn)}(a_1,r_1)
    &= \langle n | \hat{W}_{PA}(a_1,r_1) | n \rangle
    = a_1 \langle n | \hat{D}(r_1) \hat{\Pi} \hat{D}^\dagger(r_1)  | n \rangle \\
    &= a_1 \langle n | \int_{\gamma \in \mathbb{C}} \frac{d \gamma}{2 \pi} \hat{D}(r_1) \hat{D}(\gamma) \hat{D}^\dagger(r_1)  | n \rangle \\
    &= a_1 \int_{\gamma \in \mathbb{C}} \frac{d \gamma}{2 \pi} \langle n | \hat{D}(r_1) \hat{D}(\gamma) \hat{D}^\dagger(r_1)  | n \rangle \\
    &= a_1 \int_{\gamma \in \mathbb{C}} \frac{d \gamma}{2 \pi} e^{r_1(\gamma^* - \gamma)} \langle n | \hat{D}(\gamma)  | n \rangle \\
    &= a_1 \int_{\gamma \in \mathbb{C}} \frac{d \gamma}{2 \pi} e^{r_1(\gamma^* - \gamma)} e^{-\frac{1}{2}|\gamma|^2} L_n(|\gamma|^2).
\end{align}
We can use the Jacobi-Anger expansion $e^{iz\sin(\theta)}=\sum^{\infty}_{n=\infty} J_n(z)e^{in\theta}$~\cite{Abramowitz_Stegun_9} to compute the sinusoidal exponential we get from switching to cylindrical coordinates and get the following relation, 
\begin{align}
    W_{PA}^{(nn)}(a_1,r_1)
    &= a_1 \int_{0}^{\infty} d \rho \int_{0}^{2\pi} \frac{d \theta}{2 \pi} \rho e^{2ir_1 \rho \sin(\theta)} e^{-\frac{1}{2}\rho^2} L_n(\rho^2) \\
    &= a_1 \int_{0}^{\infty} d \rho \rho J_0(2r_1\rho) e^{-\frac{1}{2}\rho^2} L_n(\rho^2). 
\end{align}
The intermediate result corresponds to an eigenfunction of the Hankel transform~\cite{bateman_1954}, and we can write the diagonal witness elements as follows:
\begin{align}
    W_{PA}^{(nn)}(a_1,r_1)
    =  a_1 (-1)^n e^{-2r_1^2} L_n(4r_1^2). 
    %\overset{?}{=} \pi_{nn}(r)
\end{align}
For the general witness construction, we get the following expression:
\begin{align}
    \hat{W}_{PA}(\boldsymbol{a}, \boldsymbol{r}) &=  \sum_{j=1}^{N} \hat{W}_{PA}(a_j, r_j) \\ 
    &= \sum_{j=1}^N a_j \int_0^{2 \pi} \frac{d \theta}{2 \pi} \hat{D}(\alpha_j) \hat{\Pi} \hat{D}^\dagger(\alpha_j) \bigg|_{|\alpha|=r_j} \\
    &= \sum_{j=1}^N a_j \sum_{n=0}^\infty W_{nn}(r_j) \ket{n}\bra{n} \\
    &= \sum_{n=0}^\infty \sum_{j=1}^N a_j (-1)^n e^{-2r_j^2} L_n(4r_j^2) \ket{n}\bra{n}.
\end{align}

\subsection{Reduction of optimization parameters} \label{app:pa_witness_reduction}

As a result of the diagonal witness, we get the following simplification for the computation of the witness thresholds for the stellar rank $r^*=k$ detection. We rewrite the problem to make use of the diagonal form of the witness and reduce the number of optimization parameters by one. 
Since we optimize over the Gaussian transformation, we do the transformation $\gamma' \rightarrow\gamma$ without actually calculating the resulting shift of $\gamma'(\varphi)$, since it is an optimization parameter and the subspace of $\gamma$ is not limited by the rotation, 
\begin{align}
    \hat{O'}_{k} 
    &= \hat{\Pi}_{k} \hat{G} \hat{W}_{PA}(\boldsymbol{a}, \boldsymbol{r}) \hat{G}^\dagger \hat{\Pi}_{k} \\
    &= \hat{\Pi}_{k} \hat{D}(\gamma') \hat{S}(\xi) \hat{W}_{PA}(\boldsymbol{a}, \boldsymbol{r}) \hat{S}^\dagger(\xi) \hat{D}^\dagger(\gamma') \\
    &= \hat{\Pi}_{k} \hat{D}(\gamma') \hat{R}(\phi_s) \hat{S}(r_s) \hat{R}^\dagger(\phi_s) \hat{W}_{PA}(\boldsymbol{a}, \boldsymbol{r}) \hat{R}(\phi_s) \hat{S}^\dagger(r_s) \hat{R}^\dagger(\phi_s) \hat{D}^\dagger(\gamma') \hat{\Pi}_{k} \\
    &= \hat{\Pi}_{k} \hat{R}(\phi_s) \hat{D}(\gamma) \hat{S}(r_s) \hat{R}^\dagger(\phi_s) \hat{W}_{PA}(\boldsymbol{a}, \boldsymbol{r}) \hat{R}(\phi_s) \hat{S}^\dagger(r_s)  \hat{D}^\dagger(\gamma) \hat{R}^\dagger(\phi_s) \hat{\Pi}_{k} \\
    &= \hat{R}(\phi_s) \hat{\Pi}_{k}  \hat{D}(\gamma) \hat{S}(r_s) \hat{W}_{PA}(\boldsymbol{a}, \boldsymbol{r})  \hat{S}^\dagger(r_s)  \hat{D}^\dagger(\gamma) \hat{\Pi}_{k} \hat{R}^\dagger(\phi_s) \\
    \Rightarrow
    \hat{O}_{k} 
    &= \hat{\Pi}_{k}  \hat{D}(\gamma) \hat{S}(r_s) \hat{W}_{PA}(\boldsymbol{a}, \boldsymbol{r})  \hat{S}^\dagger(r_s)  \hat{D}^\dagger(\gamma) \hat{\Pi}_{k} \\
    &= \sum_{n=0}^\infty \Bigg( \sum_{j=0}^N W_{PA}^{(nn)}(a_j,r_j) \Bigg) \hat{\Pi}_{k}  \hat{D}(\gamma) \hat{S}(r_s) \ket{n}\bra{n}  \hat{S}^\dagger(r_s)  \hat{D}^\dagger(\gamma) \hat{\Pi}_{k}. \label{eq:reduced_opt_problem}
\end{align}
We can see in the final expression in Eq.~\eqref{eq:reduced_opt_problem} that the optimization problem of the witness thresholds reduces from four real parameters to only three real parameters for the diagonal phase-averaged witness. 

\subsection{Phase averaged optimization matrix elements} \label{app:pa_witness_integral}

This section computes the phase-average integral of the optimization matrix elements given by Eq.~\eqref{eq:pi}, and uses the generating functions of the Hermite polynomials to do so. 
The generating functions of the probabilistic Hermite polynomials are defined as 
\begin{align}
    e^{tx-t^2/2} = \sum_{n=0}^\infty \frac{t^n}{n!} \text{He}_n(x).
\end{align}
With this, we can define the integral of the generating functions as
\begin{align}
    G(u,v) &= 
    \int_0^{2\pi}\frac{d\theta}{2\pi} 
        e^{-2|z(\theta)|^2}  
        \bigg(
            \sum_{n=0}^\infty \frac{u^n}{n!} \text{He}_n\Big(2\Re\big(z(\theta)\big)\Big)
        \bigg)
        \bigg(
            \sum_{m=0}^\infty \frac{v^m}{m!} \text{He}_m\Big(2\Im\big(z(\theta)\big)\Big)
        \bigg) \\
    &=
    \int_0^{2\pi}\frac{d\theta}{2\pi} 
        \exp\Big(
            -2|z(\theta)|^2 
            + 2u\Re\big(z(\theta)\big)-\frac{u^2}{2} 
            + 2v\Im\big(z(\theta)\big)-\frac{v^2}{2}
        \Big).
\end{align}
To solve this integral, we can use the relation of $z(\theta)$ to express the integrand as an exponential of cosine functions, which are standard integrals we can solve analytically. 
First, we use the following relations to write the exponent in terms of the phase $\theta$: 
\begin{align}
    z(\theta) = \gamma + \mu e^{i\theta} - \nu e^{-i\theta}
\end{align}
\begin{align}
    |z(\theta)|^2 = |\gamma|^2 +\mu^2 +\nu^2 +2\mu\nu \underbrace{\big(\sin(\theta)^2-\cos(\theta)^2\big)}_{=-\cos(2\theta)} +2\Re(\gamma)\cos(\theta)(\mu-\nu)  +2\Im(\gamma)\sin(\theta)(\mu+\nu)
\end{align}
\begin{align}
    \Re\big(z(\theta)\big) &= \Re(\gamma) + \cos(\theta)(\mu-\nu)\\
    \Im\big(z(\theta)\big) &= \Im(\gamma) + \sin(\theta)(\mu+\nu).
\end{align}
From these relations, we get the following relation for the exponent. In the last line, we use the harmonic addition formula for the cosine, 
\begin{align}
    -2|z(\theta)|^2& + 2u\Re\big(z(\theta)\big) + 2v\Im\big(z(\theta)\big)
    \notag \\
    = 
     & 2\Re(\gamma)u + 2\Im(\gamma)v - 2(\mu^2+\nu^2) - 2|\gamma|^2 
      + 4\mu\nu\cos(2\theta) + 2(\mu-\nu)(u-2\Re(\gamma))\cos(\theta) \notag \\
     &+ 2(\mu+\nu)(v-2\Im(\gamma))\sin(\theta) \notag\\
    = 
    & 2\Re(\gamma)u + 2\Im(\gamma)v - 2(\mu^2+\nu^2) - 2|\gamma|^2 
      +\underbrace{4\mu\nu}_{=b} \cos(2\theta) \notag \\
    & +\underbrace{
          \text{sgn}\big(u-\Re(\gamma)\big) 
          \sqrt{4e^{-2r_s}\big(u-\Re(\gamma)\big)^2 + 4e^{2r_s}\big(v-\Im(\gamma)\big)^2}
        }_{=a(u,v)} \notag\\
    & \cross
    \cos\Bigg( 
      \theta+
      \underbrace{
         \arctan\bigg(e^{2r_s}\frac{u-\Re(\gamma)}{v-\Im(\gamma)}\bigg)
      }_{=\phi(u,v)} 
    \Bigg).
\end{align}
We can relate this type of integral to the modified Bessel function of the first kind with the following relation: 
\begin{align}
    e^{z\cos(\theta)} = I_0(z) + 2\sum_{k=1}^\infty I_k(z) \cos(k\theta).
\end{align}
The integral of the generating function can then be computed with
\begin{align}
    II &= \int_0^{2\pi} \frac{d\theta}{2\pi} e^{b\cos(2\theta)} e^{a\cos(\theta+\phi)} \\
    &= \int_0^{2\pi} \frac{d\theta}{2\pi} 
        \bigg(
            I_0(b) + 2\sum_{k=1}^\infty I_k(b) \cos(2k\theta)
        \bigg)
        \bigg(
            I_0(a) + 2\sum_{k=1}^\infty I_k(a) \cos\big(k(\theta+\phi)\big)
        \bigg) \\
    &= I_0(a)I_0(b) 
        + 2\sum_{k=1}^\infty I_0(a) I_k(b)  
            \underbrace{\int_0^{2\pi} \frac{d\theta}{2\pi} \cos(2k\theta)}_{=0} 
        + 2\sum_{k'=1}^\infty I_{k'}(a) I_0(b)  
            \underbrace{\int_0^{2\pi}\frac{d\theta}{2\pi} \cos\big(k'(\theta+\phi)\big)}_{=0} \notag\\
        \quad &+ 4\sum_{k,k'=1}^\infty I_{k'}(a) I_k(b)
            \int_0^{2\pi}\frac{d\theta}{2\pi} \cos(2k\theta)\cos\big(k'(\theta+\phi)\big) \\
    &= I_0(a)I_0(b) + 4\sum_{k,k'=1}^\infty I_{k'}(a) I_k(b) \frac{1}{2} \cos(k'\phi) \delta_{2k,k'} \\
    &= I_0(a)I_0(b) + 2\sum_{k=1}^\infty I_{2k}(a) I_k(b) \cos(2k\phi).
\end{align}
Now we can write out the integral over the generating function, which has essentially the same form with the addition of an exponential prefactor: 
\begin{align}
    G(u,v) = 
    \exp&\bigg( -\frac{1}{2}(u^2+v^2) +2\Re(\gamma)u +2\Im(\gamma)v -2(\mu^2+\nu^2) -2|\gamma|^2 \bigg) \notag \\
    \cross &\bigg( I_0\big(a(u,v)\big)I_0\big(b\big) + 2\sum_{k=1}^\infty I_{2k}\big(a(u,v)\big) I_k\big(b\big) \cos\big(2k\phi(u,v)\big) \bigg).
\end{align}
Using the formalism for generating functions, we can obtain the original expression by differentiation for all indices, 
\begin{align}
    \int_0^{2\pi}\frac{d\theta}{2\pi}\pi_{n,m}\big(z(\theta)\big) 
    = 
    &\sum_{p=0}^{\min(m,n)} 
        \frac{\sqrt{m!n!}}{p!(m-p)!(n-p)!} \notag\\
    & \cross \sum_{\substack{0\le s\le m-p\\0\le t\le n-p}}
            \binom{m-p}s \binom{n-p}ti^{s-t}
            \pdv{^{m+n-2p-s-t}}{u^{m+n-2p-s-t}} \pdv{^{s+t}}{v^{s+t}} G(u,v) \bigg|_{u,v=0}. 
    \label{eq:pa_opt_matrix_generating}
\end{align}
%-----------------------------------------------------------------------------------------------------------
\section{Witnessing stellar rank of Fock-states using phase-averaged witness}\label{appsec:fock7_phase_averaged}
In this section, we numerically investigate the phase-averaged witness. While the analytical treatment in the previous section established the properties of the phase-averaged construction and the reduction of the optimization problem, here we focus on its numerical implementation and performance. The main motivation is to investigate whether exploiting rotational symmetry can provide an efficient alternative to the point-based witness constructions considered in this work. We first describe the numerical implementation and its convergence, followed by an analysis of Fock states, which constitute the primary class of states targeted by the phase-averaged construction. Finally, we consider the application to phase-averaged experimentally reconstructed states.

To obtain a phase-insensitive witness for rotationally symmetric states, the few-point displaced parity matrix was numerically averaged over the phase of the selected phase-space points, with different absolute values. For a set of phase-space points $\boldsymbol{\alpha}=\{\alpha_1,...\alpha_N\}$ with corresponding radii $\boldsymbol{r}=\{r_1,...,r_N\}$ and weights $\boldsymbol{a}=\{a_1,...,a_N\}$, the phase-averaged matrix to compute the stellar-rank thresholds was calculated as
\begin{equation}
    \pi^{PA}_{nm}(\gamma, \boldsymbol{r}, r_s)
    =
    \frac{1}{N_\theta}
    \sum_{l=1}^{N_\theta}
    \pi_{nm}
    \big(
        \gamma +r_j e^{i\theta_l} \cosh(r_s) + r_j e^{-i\theta_l} \sinh(r_s)
    \big),
    \qquad
    \theta_l = \frac{2\pi l}{N_\theta},
\end{equation}
where $N_\theta$ denotes the number of uniformly spaced phase samples. The parameters $\Re(\gamma)$, $\Im(\gamma)$, and the real squeezing magnitude $r_s$ are optimized when calculating the witness thresholds. As motivated in Section~\ref{appsec:phase_averaged_witness}, the phase of the squeezing parameter is fixed to zero, while the phase-space points are sampled uniformly over the full interval $[0,2\pi)$.

For each certification level $k$, the threshold was obtained by maximizing the largest eigenvalue of the phase-averaged matrix with respect to the Gaussian transformation parameters,
\begin{equation}
    w^{PA}_k(\boldsymbol{a}, \boldsymbol{r}) =
    \underset{\gamma\in\mathbb{C}, r_s\in[0,\infty)}{\text{sup}} 
    \Bigg[ 
      \text{max eig} 
      \bigg( 
        \sum_{j=1}^N a_j \sum_{0\le m,n\le k} \pi^{PA}_{nm}(\gamma, r_j, r_s) \ket{m} \bra{n} 
      \bigg) 
    \Bigg].
    \label{eq:threshold_opt_problem_pa}
\end{equation}
The optimization was performed numerically using the Nelder--Mead algorithm and multiple randomly chosen initial points were used to reduce the dependence on the initial parameters. The real and imaginary parts of the displacement parameter were initialized uniformly in the interval $[-4,4]$, while the squeezing magnitude was initialized in the interval $[0,2]$. For the certification level $k$, the largest threshold obtained across the successful optimization runs was retained.

The phase average itself was evaluated using a finite number of uniformly spaced phase samples. Since the numerical approximation of the continuous phase integral introduces an additional discretization parameter, a separate convergence analysis was performed by increasing $N_\theta$ and monitoring the resulting threshold values. The number of phase samples used for the results presented below was chosen such that further refinement of the phase discretization produced no significant change in the calculated thresholds. This ensures that the observed behavior is not a numerical artifact caused by an insufficient number of phase samples.

\begin{figure*}[h]
\centering
\begin{minipage}{0.45\textwidth}
    \centering
    \includegraphics[width=\linewidth]{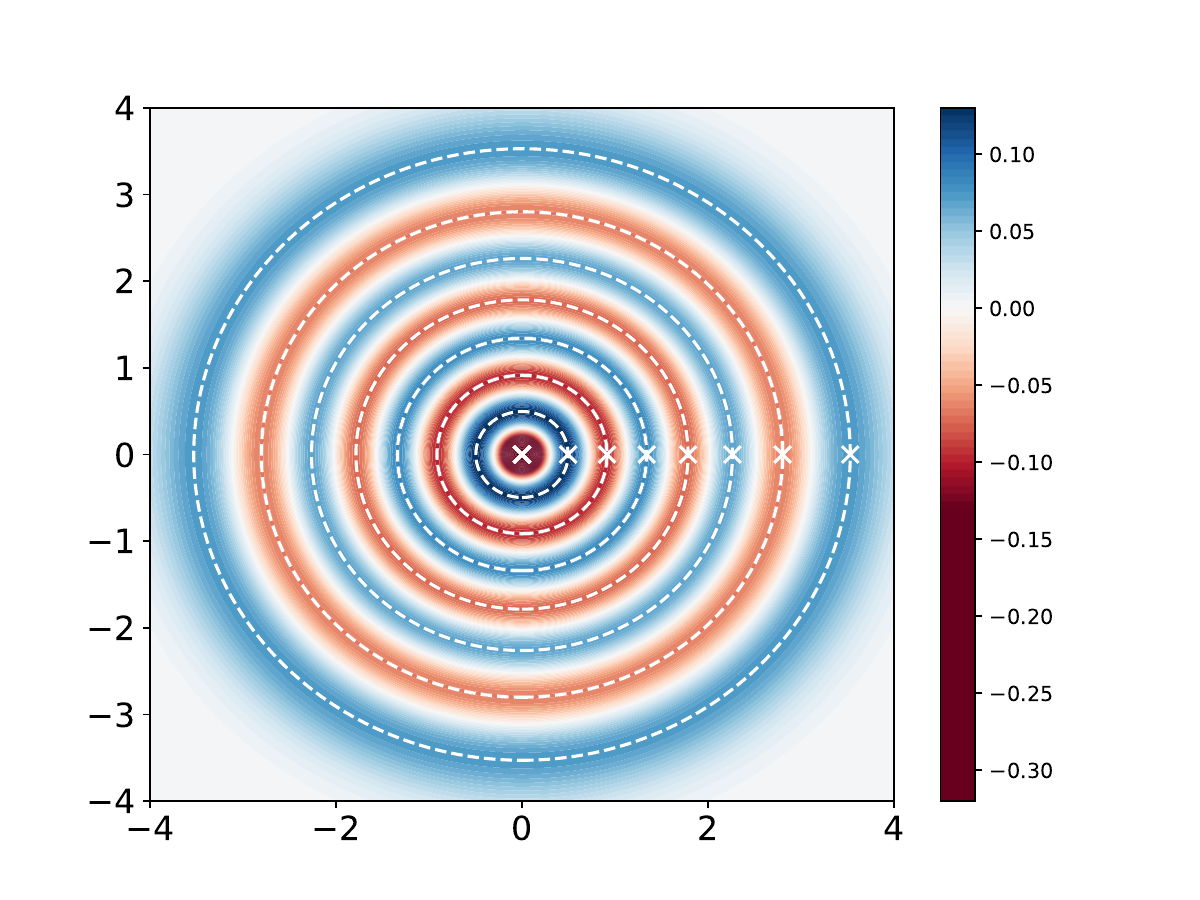}
    \caption{Wigner function of the ideal Fock state $\ket{7}$. The white crosses indicate the radii at which the phase-averaged witness is evaluated, while the dashed white circles indicate the corresponding sets of phase-space points at a fixed distance from the origin.}
    \label{fig:fock7_state-phase-averaged}
\end{minipage}
\hfill
\begin{minipage}{0.5\textwidth}
    \centering
\begin{tabular}{c|c|c|c|c}
    $w^{PA}_0$ & $w^{PA}_1$ & $w^{PA}_2$ & $w^{PA}_3$ & \\
    \cline{1-4}
    0.0286 & 0.1359 & 0.1387 & 0.1768 & \\
    \cline{1-4}
    \noalign{\vskip 8pt}
    $w^{PA}_4$ & $w^{PA}_5$ & $w^{PA}_6$ & $w^{PA}_7$ &
    $\mathrm{Tr}(\hat{W}_{PA}(\boldsymbol{a}, \boldsymbol{r})\rho)$ \\
    \hline
    0.1821 & 0.2190 & 0.2290 & 0.3545 & 0.3539 \\
    \hline
\end{tabular}
    \captionof{table}{Witness thresholds and theoretical witness expectation value for the phase-averaged witness applied to the ideal Fock state $\rho=\ket{7}\bra{7}$. An expectation value higher than a threshold value $w_k$ certifies a stellar rank $r^*> k$.}
    \label{tab:thresholds_pa_fock_7}
\end{minipage}
\end{figure*}

For the ideal Fock state $\ket{7}$, the resulting thresholds are presented in Table~\ref{tab:thresholds_pa_fock_7}, while the corresponding evaluation radii are illustrated in Figure~\ref{fig:fock7_state-phase-averaged}. The phase-averaged witness reaches an expectation value of $0.3539$, while the thresholds increase from $0.0286$ for $k=0$ to $0.3545$ for $k=7$. In particular,
[
0.3539 > 0.2290
]
for $k=6$, while
[
0.3539 < 0.3545
]
for $k=7$. The phase-averaged witness therefore certifies $r^*\ge7$, but does not certify $r^*\ge8$. This agrees with the known stellar rank $r^*=7$ of the target state and shows that the phase-averaged construction recovers its full known stellar rank.

The performance can be compared with the 22-point spiral witness considered in Table~\ref{tab:thresh_values_fock7}. For the spiral witness, the thresholds for $k=6$ and $k=7$ were $0.2532$ and $0.3069$, respectively, with an expectation value of $0.2951$. The phase-averaged witness therefore gives a substantially larger absolute witness expectation value and lower separation from the $k=7$ threshold, while still certifying the correct stellar rank of the target state. However, the two constructions are not directly equivalent in their experimental measurement requirements: the point-based witness uses a finite set of individual phase-space measurements, whereas the phase-averaged witness represents an angular average and therefore requires this average to be implemented or approximated experimentally. The phase-averaged construction should consequently not be judged solely by its threshold values.

Combined with the numerical results in section \ref{sec:phase_averaged_witness}, the numerical results here therefore demonstrate that phase averaging provides a viable symmetry-adapted construction for rotationally symmetric states. For the ideal $\ket{7}$ state, the phase-averaged witness certifies the full known stellar rank, while the experimentally reconstructed $\ket{1}$ and $\ket{2}$ states (in section \ref{sec:phase_averaged_witness}) retain their expected non-trivial stellar-rank certification. At the same time, the results do not establish that phase averaging universally improves the certified stellar rank or the experimental robustness compared with an optimized point-based witness.

A further advantage of the phase-averaged construction is the reduction of the threshold optimization problem from four to three real parameters, as established analytically in Section~\ref{appsec:phase_averaged_witness}. This reduction can be particularly useful for the non-convex numerical optimization, since more independent optimization runs can be performed for the same computational cost. At the moment, this benefit is, however, limited due to the computational cost of the phase averaging. The numerical implementation therefore provides a practical route for exploiting the rotational symmetry of Fock states while retaining the eigenvalue-based threshold formalism of the original few-point witness.

There are several directions for further investigation. First, a systematic comparison of phase-averaged and point-based witnesses for a larger set of Fock states would be required to determine how the relative threshold performance scales with the stellar rank. Second, the dependence of the phase-averaged witness on the number of sampled phases $N_\theta$ should be quantified in terms of both numerical accuracy and computational cost. Third, the experimental implementation of the angular average should be considered. Finally, the potential robustness advantage suggested by the symmetry of the phase-averaged construction could be tested directly using the displacement-error model introduced in Section~\ref{appsec:loss_displacement}.

%-------------------------------------------------------------------------------------------
\bibliographystyle{apsrmp4-2}
\bibliography{bib}
\end{document}